\PassOptionsToPackage{table,xcdraw}{xcolor}
\documentclass[journal]{IEEEtran}
\ifCLASSINFOpdf
\else
\fi

\usepackage{nicematrix}
\usepackage[utf8]{inputenc}
\DeclareUnicodeCharacter{2212}{\ensuremath{-}}
\usepackage[T1]{fontenc}
\usepackage{amsmath}
\usepackage{booktabs}
\usepackage{siunitx}
\usepackage{mathtools}
\usepackage{braket}
\usepackage{graphicx}
\graphicspath{{./figures/}}
\usepackage{amsfonts}
\usepackage[ruled,vlined,linesnumbered]{algorithm2e}
\let\oldnl\nl
\newcommand{\nonl}{\renewcommand{\nl}{\let\nl\oldnl}}
\usepackage{csquotes}
\usepackage{hhline}
\usepackage{colortbl}
\usepackage{amssymb}
\usepackage{listings}
\usepackage{color}
\usepackage{tabularx}
\usepackage{subcaption}
\usepackage{dcolumn}
\definecolor{codegreen}{rgb}{0,0.6,0}
\definecolor{codegray}{rgb}{0.5,0.5,0.5}
\definecolor{codeamethyst}{rgb}{0.6, 0.4, 0.8}
\definecolor{ao}{rgb}{0.0, 0.0, 1.0}
\definecolor{azure(colorwheel)}{rgb}{0.0, 0.5, 1.0}
\definecolor{backcolour}{rgb}{0.95,0.95,0.92}
\usepackage{titlesec}

\usepackage{hyperref}
\hypersetup{
    colorlinks=true,
    linktoc=all,
    linkcolor=black,
    citecolor=blue,
    allcolors=ao
}
\usepackage{longtable}
\usepackage{braket}
\newcolumntype{C}{>{\centering\arraybackslash}X}
\SetAlFnt{\small}
\usepackage[font=small]{caption}
\begin{document}

\title{Analysis of The Vehicle Routing Problem Solved via Hybrid Quantum Algorithms in Presence of Noisy Channels}

\author{\IEEEauthorblockN{1\textsuperscript{st}Nishikanta Mohanty,}
\IEEEauthorblockA{\textit{Centre for Quantum Software and Information, } \\
\textit{University of Technology Sydney, } Ultimo, Sydney 2007, NSW, Australia \\ Nishikanta.M.Mohanty@student.uts.edu.au}\\
\and
\IEEEauthorblockN{2\textsuperscript{nd} Bikash~K.~Behera, }
\IEEEauthorblockA{\textit{Bikash's Quantum (OPC) Pvt. Ltd., } Mohanpur 741246, WB, India\\
bikas.riki@gmail.com}\\
\and
\IEEEauthorblockN{3\textsuperscript{rd} Christopher Ferrie, }
\IEEEauthorblockA{\textit{Centre for Quantum Software and Information,} \\
\textit{University of Technology Sydney,} Ultimo, Sydney 2007, NSW, Australia \\
Christopher.Ferrie@uts.edu.au}}


\maketitle

\begin{abstract}
The vehicle routing problem (VRP) is an NP-hard optimization problem that has been an interest of research for decades in science and industry. The objective is to plan routes of vehicles to deliver goods to a fixed number of customers with optimal efficiency. Classical tools and methods provide good approximations to reach the optimal global solution. Quantum computing and quantum machine learning provide a new approach to solving combinatorial optimization of problems faster due to inherent speedups of quantum effects. Many solutions of VRP are offered across different quantum computing platforms using hybrid algorithms such as quantum approximate optimization algorithm and quadratic unconstrained binary optimization. In this work, we build a basic VRP solver for 3 and 4 cities using the variational quantum eigensolver on a fixed ansatz. The work is further extended to evaluate the robustness of the solution in several examples of noisy quantum channels. We find that the performance of the quantum algorithm depends heavily on what noise model is used. In general, noise is detrimental, but not equally so among different noise sources.
\end{abstract}


\maketitle

\section{Introduction}
\label{sec:introduction}
Quantum computers, the next generation of computing technology, are expected to solve complex optimisation problems much faster than their traditional counterparts. As parallelism is quantum computing's most notable benefit \cite{montanaro_quantum_2016,jordan_httpsquantumalgorithmzooorg_nodate}, it is only natural to turn to quantum computing to speed up calculations in complicated optimisation problems (such as those described by quantum approximate optimisation algorithm (QAOA) \cite{farhi_quantum_2014}, adiabatic computation (AC) \cite{farhi_quantum_2000}, Grover's algorithm \cite{grover_fast_1996}, and others). When applied to a multidimensional problem, classical optimisation techniques in machine learning (ML) can take a long time to calculate global optimum and consume a lot of CPU, and GPU power \cite{dasari_solving_2020}. In higher dimensional problem spaces, classical algorithms have been shown to be less effective in general \cite{nac_chapter3_quantum_algo}. This is due to the fact that NP-hard optimisation tasks are often assigned to ML algorithms. \cite{dasari_solving_2020}.

VRP comes under the category of routing problems that try to address multiple issues related to fleet management \cite{ harwood_formulating_2021}. The objective is always to optimize vehicle movement to minimize the cost or maximize the profit. Notwithstanding the difficulty in delivering quick and dependable solutions to the computationally hard VRP problem, several precise and heuristic techniques have been developed for solving it \cite{harwood_formulating_2021,srinivasan_efficient_2018}. Describing the VRP in its simplest form, a single vehicle is tasked to deliver goods at multiple customer locations; also, the vehicle needs to return to pick up additional items when it runs out of goods \cite{feld_hybrid_2019}. The goal is to minimize the cost of service by finding the best feasible combination of routes that begin and terminate at a central location (the depot) while maximizing the reward (often the inverse of the total distance traveled or the mean service time). This problem is computationally difficult to solve, even with only a few hundred customer nodes, and is classified as an NP-hard problem \cite{dasari_solving_2020, nazari_reinforcement_2018}.

Any VRP $(n, k)$ involves  $(n - 1)$ locations with k vehicles and a depot $D$ \cite{utkarsh_solving_2020, harwood_formulating_2021}. Its solution is a set of routes in which all $k$ vehicles begin and terminate at the given depot $D$, ensuring that each place is visited only once. The route with the shortest sum total of distance traveled by $k$ vehicles is the ideal one. For a long time, VRP is studied as an extension of the classical traveling salesman problem \cite{papalitsas_qubo_2019, srinivasan_efficient_2018}, where now a group of $k$ salesmen has to service collectively $(n - 1)$ locations, such that each location is serviced exactly once \cite{harwood_formulating_2021}. Constraints such as vehicle capacity or restricted covering time often complicate the VRP issue in practical settings. As a result, a plethora of conventional and quantum approaches have been presented in an effort to effectively solve the problem. Current quantum approaches for solving optimization problems include QAOA \cite{farhi_quantum_2014}, Quadratic unconstrained binary optimization (QUBO) \cite{glover_quantum_2020,kochenberger_unconstrained_2014}, quantum annealing  \cite{irie_quantum_2019, crispin_quantum_2013, fujitsu_annealer_2019}, and Variational quantum eigensolver(VQE) \cite{utkarsh_solving_2020}, which we will define in detail later.

In this work, we study the VRP in a different light. Here we explore adding controlled noise to an adapted quantum solution to determine if it improves or degrades the overall results. Recent works in QAOA \cite{campos_training_2021,wang_noise-induced_2021,Marshall_2020, lavrijsen_classical_2020} and VQE algorithms \cite{cerezo_variational_2021} studied the generic effects of noise in these hybrid algorithms. Our work complements these results by analyzing the effects of noise in a detailed gate-based simulation of an algorithm to solve VRP. We analyze the effect of various noise channels on an existing, yet variable, ansatz developed as a solution to VRP. We apply amplitude damping, bit-flip, phase-flip, bit-phase-flip, and depolarising noise channels to VRP circuits, analyze the effects, and consolidate our findings.

The paper is organized as follows. Sec. \ref{VRP} discusses fundamental mathematical concepts such as combinatorial optimization, adiabatic computation, QAOA, and the Ising model. Sec. \ref{Modelling} discusses the formulation of VRP using the concepts discussed in the previous Section. Sec. \ref{Analysis And Circuit Building} covers the basic building blocks of circuits to solve VRP. Sec. \ref{VQE Simulation} covers building an ansatz for VRP. Finally, Sec. \ref{Noise Model Simulation of VQE} covers the effects of applying noise models on the VRP circuit. Then Sec. \ref{Inferences from Simulation} presents the observations from the simulation results. In Sec. \ref{Discussion and Conclusion}, we summarize the effects of various noise models on the VRP circuit and future directions of research.

\section{Mathematical Background} \label{VRP}
The fundamental concepts used to solve routing problems involve techniques and procedures from the field of combinatorial optimization. This is followed by converting the mathematical models to a quantum equivalent mathematical model for formulating the objective function. The solution of the objective function is often achieved by maximization or minimization of the function. In this section, we outline the key concepts. 

\subsection{Combinatorial Optimization} 
A classical combinatorial optimization (CO) problem is finding an optimal object from a finite set of objects. Exhaustive search is impractical in finding the optimal object due to the potentially high number of objects. Mathematically defining, if $s$ is a string in some set $S$ and $m$ are some clauses, where $s \geq m$, we have a maximization or minimization problem,  known as a CO problem. Each clause expects a string parameter and returns a corresponding value\cite{guerrero_solving_2020}. It is the sum over the $m$ clauses that constitutes the string's total cost function. If we refer to the input string as $z$ and clauses as $C_\alpha$, we can write the total cost function as

\begin{eqnarray}
C(z)=\sum^m_{\alpha{=1}}{C_{\alpha}\left(z\right)}.
\end{eqnarray}

The objective is to identify $\overline{z}\in S$ such that $C\left(\overline{z}\right)\ge C_{\alpha}(z)$ for all $z \in S$ (or, in the case of minimization, $ C\left(\overline{z}\right){\le }C_{\alpha}(z)$ for all $z\in S$). Here $\overline{z}$ is not required to be unique and  if  $z$ satisfies clause $\alpha$; $C_\alpha(z)=1$  else it's $0$. $z$ can be written as $z{=}z_0z_{{1}}z_{{2}}{\dots \dots }z_{n{-}{1}}$ for $z_i{\ }{\in }{\{}{0,1}{\}}$. Also, considering only maximization problems, the minimization problems can be studied as $C^{{'}}_{\alpha}{(}z{)=1-}C_{\alpha}{(}z{)}$

\begin{eqnarray}
C^{{'}}({z})&=&\sum^{m-1}_{k=0}{\ }C^{{'}}_{\alpha}({z})=\sum^{m-1}_{k=0}{\ }\left(1-C_{\alpha}({z})\right),\nonumber\\
&=&m-\sum^{m-1}_{k=0}{\ }C_{\alpha}({z})=m-C({z}).
\end{eqnarray}

\subsection{ Adiabatic Quantum Computation}
Adiabatic quantum computation (AQC) is a theoretical framework of a quantum computer \cite{farhi_quantum_2000, albash_adiabatic_2018}. The adiabatic theorem asserts that if the change to the Hamiltonian is sufficiently gradual, the system remains in the ground state of the given Hamiltonian\cite{grant_adiabatic_2020}. The Hamiltonian is an energy operator of a system. In AQC, there are two Hamiltonians: the driver Hamiltonian ($H_d$) and the problem Hamiltonian ($H_p$). The driver Hamiltonian ($H_d$) is the energy operator whose ground state is easy to prepare, whereas the problem Hamiltonian ($H_p$) is the energy operator whose ground state is obtained after evolution  \cite{farhi_quantum_2000}. Interpolation times are proportional to the energy gap between the two lowest states of the Hamiltonian being used. 

The procedure begins with an easy-to-prepare ground state (i.e., the ground state of ${(}H_d{)}$) and ends (ideally) with the ground state of ${(}H_p{)}$, which is, in general, not directly characterizable. Mathematically, constitute function $s{(}t{)}$ on $\left[0,T\right]$ where $s\left(0\right){=0\ }$ and $s\left(T\right){=1}$. $T$ is the value of time set high enough for the adiabatic theorem to hold. We define the Hamiltonian, $H(t)=(1-s(t))H_D+s(t)H_P$. According to the adiabatic theorem, a system maintains its initial state of $H{(}t{)}$ across the whole interval $[0, T]$, provided a suitable $s(t)$; hence, the system is in the initial ground state ${(}H_d{)}$ at time $t = 0$, and it will evolve into the intended ground state ${(}H_p{)}$ at time $t = T$. 
In general, it is challenging to assess the integral describing the temporal evolution under this time-dependent Hamiltonian \cite{hoque_quantum_nodate}:

\begin{eqnarray}
U(t)=\tau \operatorname{exp}\left\{\frac{-i}{\hbar}\int^t_0{\ }H(T)dT\right\}.
\label{}
\end{eqnarray}

It is possible to assess this Hamiltonian using Trotterization methods \cite{sun_adiabatic_2018}. We divide $U(T)$ into intervals of ${\delta}t$ small enough such that the Hamiltonian is almost constant over them. This permits us to use the much more streamlined formula for the Hamiltonian that is independent of time. Assuming $U(b, a)$ is the time evolution from instant a to instant b.

\begin{eqnarray}
U(T,0)&=&U(T,T-\delta t)U(T-\delta t,T-2\delta t)\cdots U(\delta t,0), \nonumber\\ 
&=&\prod^p_{j{=1}}{{\ }}U{(}j{\delta}t{,(}j{-}{1)}{\delta}t{)}, \nonumber\\ 
&\approx& \prod^p_{j{=1}}{{\ }}e^{{-}iH{(}j{\delta}t{)}{\delta}t}.
\end{eqnarray}

Where the approximation gets better as $p$ gets larger (or, in other words, as ${\delta}t$ gets smaller), and where ${\delta}t$ is measured in $\hbar$. Now using the approximation $e^{i{(}A{+}B{)}x}{=}e^{iAx}e^{iBx}{+}\mathcal{O}\left(x^{{2}}\right)$ and adding Hamiltonian $H{(}j{\delta}t{)=(1-}s{(}j{\delta}t{))}H_D{+}s{(}j{\delta}t{)}H_P$ the integral $U(t)$ becomes,

\begin{eqnarray} 
U(T,0)\approx \prod^p_{j=1} e^{-i(1-s(j\delta t))H_D\delta t} e^{ -is(j\delta t)H_P\delta t}.
\end{eqnarray}

 AQC may be approximated by allowing the system to develop under $H_P$ for a small $s{(}j{\delta}t{)}{\delta}t$ and then $H_D$ for a small ${\ (1-}s\left(j{\delta}t\right){)}{\delta}t$ , and unitaries can be derived for these operations using $U{=}e^{{-}i\alpha H{\delta}t}$. Here,  $\alpha$ is an integer in the range $[0,1]$, and this includes the scaling resulting from $s{(}j{\delta}t{)}$. AQC forms the theoretical basis of the variational quantum algorithm QAOA, which is discussed briefly in the next section.

\subsection{QAOA} \label{QAOA}
Quantum Approximate Optimization Algorithm (QAOA) is a variational algorithm proposed by Farhi \emph{et al.} in 2014 \cite{farhi_quantum_2000, farhi_quantum_2014}. This algorithm relies on the framework of adiabatic quantum computation. Because of its use of both conventional and quantum methods, this algorithm is considered as hybrid algorithm. In the previous section, quantum adiabatic computation drove the system from the eigenstate of driver Hamiltonian to that of the eigenstate of problem Hamiltonian. 

In the context of optimization problems, the problem Hamiltonian can be written as,

\begin{eqnarray}
C|z\rangle =\sum^{m}_{\alpha=1}C_{\alpha(z)}|z\rangle.
\end{eqnarray}

The maximum energy eigenstate of C solves the combinatorial optimization problem. Similarly, for driver Hamiltonian we use,

\begin{eqnarray}
B=\sum^n_{j=1}{\ }{\sigma}^x_j,
\end{eqnarray}

where ${\sigma}^x_j$ represents the Pauli operator ${\sigma}^x$  on the bit $z_j$ . $B$ is also known as the mixing operator. Let us also define $U_c\left(\gamma\right){=}e^{{-}i\gamma C_{\alpha}}$ and $U_B\left(\beta\right){=}e^{{-}i\beta{B}}$ which lets the system evolve under $C$ for some $\gamma$ amount of time and under $B$ for some $\beta$ amount of time, respectively. QAOA then constructs a state, 

\begin{equation}
|\boldsymbol{\beta},\boldsymbol{\gamma}\rangle =e^{-i{\beta}_pB}e^{-i{\gamma}_p{C}}\dots e^{-i{\beta}_2B}e^{-i{\gamma}_2{C}}e^{-i{\beta}_1{B}}e^{-i{\gamma}_1{C}}|s\rangle,
\end{equation}
where $|s\rangle $ is a superposition state of all input qubits. The expectation value of the cost function $\sum^m_{\alpha{=1}}{{\langle }\beta{,}\gamma~{|C_\alpha}~{|}\beta{,}\gamma~{\rangle }}$ gives the solution or the approximate solution of the problem after simplex or gradient-based optimization starting with some initial set of parameters \cite{zhou_quantum_2020}.

\subsection{Ising Model}
The Ising model of ferromagnetism is a well-established mathematical model used extensively in the field of statistical mechanics \cite{singh_Ising_2020,RevModPhys.39.883}. There are two possible states for the magnetic dipole moments of atomic "spins" ($+1$ and $1$), each of which is represented by a discrete variable in the model. Each spin is able to communicate with its neighbors because of how they are organised in a graph, usually a lattice (where the local structure regularly repeats in all directions). The system tends towards the lowest energy state when neighboring spins agree, but heat interrupts this tendency, allowing for the emergence of alternate structural phases. The model serves as a simplification of reality that may be used to spot phase transitions \cite{lucas_Ising_2014}. Using the following Hamiltonian, we can describe the sum of the spin energies:

\begin{eqnarray}
H_c=-\sum_{\left\langle i,j\right\rangle }{\ }J_{ij}{{\sigma}}_{{i}}{{\sigma}}_{{j}}-h\sum {{\sigma}}_{{i}},
\end{eqnarray}

where $J_{ij}$ represents the interaction between $i$ and $j$, which are adjacent spins, and $h$ represents an external magnetic field. If $J$ is positive, the ground state at $h=0$ is a ferromagnet. If $J$ is negative, the ground state at $h=0$ is an anti-ferromagnet for a bipartite lattice. Hence for simplification and in the context of this document, we can write the Hamiltonian as

\begin{eqnarray}
H_c=-\sum_{\langle i,j\rangle }{\ }J_{ij}{\sigma}_{i}^z{\sigma}_{j}^z-\sum h_i{\sigma}_{i}^x.
\end{eqnarray}

Here ${\sigma}_z$ and ${\sigma}_x$ represent Pauli $z$ and $x$ operator. For simplification, we can consider the following conditions to be ferromagnetic ($J_{ij}>0$), $h=0$ assuming no external influence on the spin. Thus we can rewrite the Hamiltonian as follows,

\begin{eqnarray}
H_c=-\sum_{\langle i,j\rangle }{\ }J_{ij}{\sigma}_{i}^z{\sigma}_{j}^z=-\sum_{\langle i,j\rangle }{\ }\sigma_{i}^z\sigma_{j}^z.
\end{eqnarray}

\subsection{VQE}
Variational Quantum Eigensolver (VQE) is a hybrid quantum-classical technique used to determine the eigenvalue of a large matrix or Hamiltonian $H$ \cite{peruzzo_variational_2014}. The basic objective of this method is to search for a trial qubit state of a wave function $\ket{\psi(\vec\theta)}$ which is dependent on a parameter set $\vec\theta = \theta_1,\theta_2,\dots$ also called as the variational parameters. By quantum theory the expectation of an observable or Hamiltonian $H$ in a state $\ket{\psi(\vec\theta)}$ can be expressed as,

\begin{eqnarray}
E(\vec{\theta})= \bra{\psi(\vec\theta)} H \ket{\psi(\vec\theta)}.
\label{Eq. Expectation}
\end{eqnarray} 

By spectral decomposition  $H$ can be written as 

\begin{eqnarray}
H= \sum_{i=1}^n \lambda_i \ket{\psi}_i \bra{\psi}_i.\nonumber\\
\end{eqnarray}

Where $\lambda_i$ and ${\ket\psi}_i$ are the eigenvalues of matrix $H$. Also the eigenstates of $H$ are orthogonal so $\left\langle\psi_{i} \mid \psi_{j}\right\rangle=0$ if $i \neq j$. The wave function $\ket{\psi(\vec\theta)}$ is represented as the superposition of eigenstates,

\begin{eqnarray}
\ket{\psi(\vec\theta)}= \sum_{i=1}^n \alpha_{i}(\vec\theta)\ket\psi_{i}.
\end{eqnarray}

Thus the expectation becomes 

\begin{eqnarray}
E(\vec\theta )  &=& \sum_{i=1}^n |\alpha_{i}(\vec\theta ) |^2 \lambda_{i}.
\end{eqnarray}

Clearly, $ E(\vec\theta) \geq \lambda_{\min }$.
So in VQE algorithm, we vary the parameters $\vec{\theta}=\theta_{1}, \theta_{2}, \ldots$ until $E(\vec{\theta})$ is minimized. This characteristic of VQE is important for addressing combinatorial optimization problems in which a parameterized circuit is used to construct the trial state of the algorithm, and $E(\vec{\theta})$ is the cost function, which is the expected value of the Hamiltonian in the trial state. On the assumption that the ansatz or parameterized quantum circuit may describe the ground state of the system, the ground state of the target Hamiltonian may be obtained by iterative minimisation of the cost function.. At each optimization stage, the optimization procedure employs a classical optimizer that employs a quantum computer to analyze the cost function and calculate its gradient.

\section{Modelling VRP in Quantum} \label{Modelling}
To find a solution to the vehicle routing problem, we can map the cost function to an Ising Hamiltonian $H_c$ \cite{lucas_Ising_2014}. The minimization of Ising Hamiltonian $H_c$ gives the solution to the problem. To begin, let us consider an arbitrary connected graph of $n$ vertices and a binary decision variable $x_{ij}$ who has a value $1$ if there exists an edge between $i$ and $j$ for edge weight $w_{ij}>0$ else; the value is $0$. To represent the VRP problem, we need $n \times (n-1)$ decision variables. For every edge from $i\rightarrow j$, we define two sets of nodes $source\left[i\right]$ and $target[j]$. The set $source\left[i\right]$ contains the nodes $j$ to which $i$ sends an edge $j\ \epsilon\ source [i]$. The set $target\left[j\right]$ contains the nodes $i$ to which $i$ sends an edge $i\ \epsilon\ target[j]$. We define VRP as follows,

\begin{eqnarray} 
VRP(n,k)=\mathop{min}_{{\left\{x_{ij}\right\}}_{i\to j}\in \{0,1\}}\ \sum_{i\to j}{\ }w_{ij}x_{ij},
\end{eqnarray}

where $k$ is the number of vehicles and $n$ is the total number of locations. Considering the starting location as $0^{th}$ location or Depot $D$, we have $n-1$ locations for vehicles to travel. This is subject to the following constraints,

\begin{eqnarray}
\sum_{j\in ~{source}~[i]}{\ }x_{ij}&=&1,{\forall }i\in \{1,\dots ,n-1\} , \nonumber\\
\sum_{j\in ~{target}~[i]}{\ }x_{ji}&=&1,{\forall }i\in \{1,\dots ,n-1\} , \nonumber\\
\sum_{{j}{\in }~{source}~{[}{0}{]}}{{\ }}{x}_{0j}&=& k, \nonumber\\
\sum_{j\in ~{target}~[0]}{\ }x_{j0}&=&k \nonumber \\
u_i-u_j+Q x_{i j} &\leq& Q-q_j, \forall i \sim j, i, j \neq 0, \nonumber\\
 q_i \leq u_i &\leq& Q, \forall i, i \neq 0.
\end{eqnarray}

The first two constraints impose the restriction that the delivering vehicle must visit each node only once. The middle two constraints enforce the restriction that the vehicle must return to the depot after delivering the goods. The last two constraints impose the sub-tour elimination conditions and are bound on $u_i$, with $Q>q_j>0$, and $u_i,Q, q_i \in \mathbb{R}$. 

For the VRP equation, we can form the Hamiltonian of VRP as follows \cite{utkarsh_solving_2020}, 

\begin{eqnarray}
H_{VRP}&=&H_A+H_B+H_C+H_D+H_E, \nonumber\\
H_A&=&~\sum_{i~\to j}{w_{ij}x_{ij}} ,\nonumber\\
H_B&=&A\sum_{i\in 1,\dots ,n-1}{\ }{\left(1-\sum_{j\in ~{source}~[i]}{\ }x_{ij}\right)}^2, \nonumber\\
H_C&=&A\sum_{i\in 1,\dots ,n-1}{\ }{\left(1-\sum_{j\in ~{target}[i]}{\ }x_{ji}\right)}^2, \nonumber\\ 
H_D&=&A{\left(k-\sum_{j\in ~{source}[0]}{\ }x_{0j}\right)}^2, \nonumber\\
H_E&=&A{\left(k-\sum_{j\in ~{target}[0]}{\ }x_{j0}\right)}^2 ,
\label{Eq. VRP}
\end{eqnarray}
where $A > 0$ is a constant. 

The set of all binary decision variables $x_{ij}$ can be represented in vector form as,

\begin{eqnarray}
\overrightarrow{\boldsymbol{{x}}}={\left[x_{(0,1)},x_{(0,2)},\dots x_{(1,0)},x_{(1,2)},\dots x_{(n-1,n-2)}\right]}^{\boldsymbol{{T}}}.
\end{eqnarray}

Using the above vector, we can define two additional vectors for each node, 

\begin{eqnarray}
\overrightarrow{z}_{S\left[i\right]}=\{x_{ij}=1,\ x_{kj}=0\ ,\ k\neq i\ ,\ \ \forall j,k\ \in \{0,\dots ,n-1\} , \nonumber\\ 
\overrightarrow{z}_{T\left[i\right]}=\{x_{ji}=1,\ x_{jk}=0\ ,\ k\neq i\ ,\ \ \forall j,k\ \in \{0,\dots ,n-1\} .\nonumber\\
\end{eqnarray}

The above vectors will assist in the formulation of the QUBO model of VRP. For a linked graph $G=(N,V)$, the QUBO model \cite{date_efficiently_2019, glover_quantum_2020,kochenberger_unconstrained_2014,guerreschi_solving_2021} is defined as,

\begin{eqnarray}
f(x)_{\text QUBO}=\mathop{min}_{x\in \{0,1\}(N\times V)}x^TQx+g^Tx+c ,
\label{Eq. QUBO}
\end{eqnarray}

where $Q$ is a quadratic coefficient of the edge weights, $g $ is a linear coefficient of the node weights, and $c$ is a constant. Here, $\mathbb{J}$ is the matrix of all ones, $\mathbb{I}$ is the identity matrix, and $e_0={\left[1,0,\dots ..,0\right]}^T$. Hence for QUBO formulation of Eq. \eqref{Eq. VRP} the coefficients are given as follows,

\begin{eqnarray}
Q&=&A\left[\left[z_{T[0]}, \ldots, z_{T[n-1]}\right]^{T}\left[z_{T[0]}, \ldots, z_{T[n-1]}\right]\right. \nonumber\\
&&\left.+\left(\mathbb{I}_{n} \otimes \mathbb{J}(n-1, n-1)\right)\right], \nonumber\\
g&=&W-2 A k\left(\left(e_{0} \otimes \mathbb{J}_{n-1}\right)+\left[z_{T[0]}\right]^{T}\right), \nonumber\\
\quad&&+2 A\left(\mathbb{J}_{n} \otimes \mathbb{J}_{n-1}\right), \nonumber\\
c&=&2 A(n-1)+2 A k^{2} .
\end{eqnarray}

The binary decision variable $x_{ij}$ is transformed to spin variable $s_{ij}  \in \left\{-1,1\right\}$ as  $x_{ij}=(s_{ij}+1)/2$.

From the above Eqs, we can expand Eq. \eqref{Eq. QUBO} to form the Ising Hamiltonian of VRP, \cite{glover_quantum_2020}

\begin{eqnarray}
H_{Ising}=-\sum_i{\ }\sum_{i<j}{\ }J_{ij}s_is_j-\sum_i{\ }h_is_i+d . 
\end{eqnarray}

Here, the terms $J_{ij}, h_i$ and $d$ are defined as follows,

\begin{eqnarray}
J_{ij}&=&\ -\frac{Q_{ij}}{2},\ \forall \ i<j  ,\nonumber\\
h_i&=&\frac{g_i}{2}+\sum{\frac{Q_{ij}}{4}+\ \sum{\frac{Q_{ji}}{4}\ }\ }, \nonumber\\
d&=&c+\sum_i{\ }\frac{g_i}{2}+\sum_i{\ }\sum_j{\ }\frac{Q_{ij}}{4} .
\end{eqnarray}

\begin{figure}[!ht]
\centering
\begin{subfigure} {0.5\linewidth}
\includegraphics[width=\linewidth]{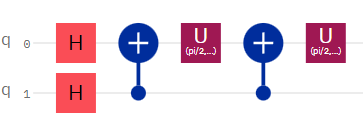} 
\caption{}
\label{qdctc_Fig3a}
\end{subfigure}
\hfill
\begin{subfigure} {0.5\linewidth}
\includegraphics[width=\linewidth]{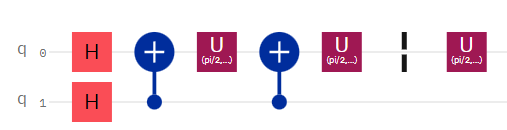} 
\caption{}
\label{qdctc_Fig3b}
\end{subfigure}
\hfill
\caption{(a) Sample circuit showing gate selections for ${H}_{\mathrm{cost}}$. (b) Sample circuit showing gate selections with additional $U$ gate after barrier for ${H}_{\text{mixer}}$. Note: The sample circuits displayed in the figures represent the building blocks of the actual circuit and do not represent actual angles that are obtained as a solution of VRP using VQE.}
\label{qdctc_Fig3}
\end{figure}

\section{Analysis And Circuit Building} \label{Analysis And Circuit Building}

In this section, we create a gate-based circuit to realize the above formulation using the IBM gate model, which we have implemented using the Qiskit framework \cite{Qiskit}. For any arbitrary VRP problem using qubits, we begin with the state of $\ket{+}^{\otimes n(n-1)}$ the ground state of $H_{\text{mixer}}$ by applying the Hadamard to all qubits initialized as zero states, and we prepare the following state.
\begin{eqnarray}
\ket{\beta,\gamma}&=&e^{-iH_{\text{mixer}}\beta_{p}}e^{-iH_{cost}\gamma_{p}}...\nonumber\\
&&...e^{-iH_{\text{mixer}}\beta_{0}}e^{-iH_{cost}\gamma_{0}}\ket{+}^{n\otimes(n-1)}.
\label{eq. State}
\end{eqnarray}

The energy $E$ of the state $\ket{\beta,\gamma}$ is calculated by the expectation of $H_{cost}$ from  Eq. \eqref{Eq. Expectation}. Once again, the $H_{cost}$ term may be expressed in terms of Pauli operators using the Ising model, as

\begin{eqnarray}
{H}_{\mathrm{cost}}=-\sum_{i} \sum_{i<j} J_{ij}\sigma_i^z\sigma_j^z-\sum_{i} h_i\sigma_i^z-d.
\end{eqnarray}

Thus for a single term of state in $\ket{\beta,\gamma}$ as $\beta_0,\gamma_0$, the expression reads,

\begin{math} 
e^{-i{H}_{mixer}\beta_0}e^{-i{H}_{cost}\gamma_0}. 
\end{math} 
The first term ${H}_{\mathrm{cost\ }}$ can be expanded to following,

\begin{eqnarray}
{e}^{iJ_{ij}\gamma_0\sigma_i\sigma_j}&=&\cos J_{ij}\gamma_0I+i\ \sin J_{ij}\gamma_0\sigma_i\sigma_j,\nonumber\\
&=& \left[\begin{matrix}{e}^{i{J}_{ij}{\gamma}_{0}} &0&0&0\\ 0&{e}^{-i{J}_{ij}{\gamma}_{0}}&0&0\\0&0&{e}^{-i{J}_{ij}{\gamma}_{0}}&0\\0&0&0&{e}^{i{J}_{ij}{\gamma}_{0}}\end{matrix}\right] \nonumber\\
&=& M.
\end{eqnarray}

Applying $CNOT$ gate on before and after the above matrix `$M$' we can swap the diagonal elements,

\begin{eqnarray}
CNOT(M)CNOT=\left[\begin{matrix}{e}^{i{J}_{ij}{\gamma}_{0}} &0&0&0\\ 0&{e}^{-i{J}_{ij}{\gamma}_{0}}&0&0\\0&0&{e}^{i{J}_{ij}{\gamma}_{0}}&0\\0&0&0&{e}^{-i{J}_{ij}{\gamma}_{0}}\end{matrix}\right].\nonumber\\
\end{eqnarray}

Observing the upper and lower blocks of the matrix, we can rewrite,

\begin{eqnarray}
\left[\begin{matrix}1&0\\0&1\end{matrix}\right]\otimes\left[\begin{matrix}{e}^{i{J}_{ij}{\gamma}_{0}}&0\\0&{e}^{-i{J}_{ij}{\gamma}_{0}}\end{matrix}\right] = I\otimes {e}^{i{J}_{ij}{\gamma}_{0}}\left[\begin{matrix}1&0\\0&{e}^{-2i{J}_{ij}{\gamma}_{0}}\end{matrix}\right].\nonumber\\
\label{Eq29}
\end{eqnarray}

$\left[\begin{matrix}1&0\\0&{e}^{-2i{J}_{ij}{\gamma}_{0}}\end{matrix}\right]$ is a phase gate. Looking at the $2nd$ term of ${H}_{\mathrm{cost}}$ we get,

\begin{eqnarray}
H_{cost}&=&\sum_{i} h_i\sigma_{i}^{z}, \nonumber\\
e^{ih_{i}\gamma_0\sigma_i^z}&=&\cos{h_i\gamma_0}I+i\sin{h_i\gamma_0}\sigma_i^z,\nonumber\\
&=&\cos{h_i\gamma_0}\left[\begin{matrix}1&0\\0&1\\\end{matrix}\right]+i\sin{h_i\gamma_0}\left[\begin{matrix}1&0\\0&-1\\\end{matrix}\right],\nonumber\\
&=& \left[\begin{matrix}e^{ih_i\gamma_0}\ &0\\0&e^{-ih_i\gamma_0}\\\end{matrix}\right]\nonumber\\
&=& e^{ih_i\gamma_0}\left[\begin{matrix}1\ &0\\0&e^{-2ih_i\gamma_0}\\\end{matrix}\right].
\label{Eq30}
\end{eqnarray}

Fig. \ref{qdctc_Fig3a} depicts the basic circuit with two qubits along with gate selections for ${H}_{\mathrm{cost}}$. 

Similarly $H_{\text{mixer}}$  can be derived as follows 
\begin{eqnarray}
&& H_{\text {mixer }}=-\sum_i \sigma_x
\end{eqnarray}

Considering a  single term of $H_{\text{mixer}}$ and taking the unitary,
\begin{eqnarray}
e^{-i H_{\text{mixer}} \beta_0}&=&e^{-i\left(-\sigma_x\right) \beta_0} \nonumber\\
&=&e^{i \sigma_x \beta 0} \nonumber\\
&=&\cos{\beta_0} \mathbb{I}+i \sin{\beta_0} \sigma_x\nonumber\\
&=&\left[\begin{array}{cc}
\cos \beta_0 & i \sin \beta_0 \\
i \sin \beta_0 & \cos \beta_0
\end{array}\right]
\label{Eq31}
\end{eqnarray}

The IBMQ $U$ gate is defined as follows. 

\begin{equation}
U=\left[\begin{array}{cc}
\cos \theta / 2 & -e^{i \lambda} \sin \theta / 2 \\
e^{i \phi} \sin \theta / 2 & e^{i(\lambda+\phi)} \cos \theta / 2\\

\end{array}\right]
\label{Eq32}
\end{equation}

Comparing Eqs. \eqref{Eq29}, \eqref{Eq30}, \eqref{Eq31} and \eqref{Eq32}, we can establish relation of circuit parameters with $\gamma,\beta$ to $U$ gate which and will form building blocks of circuit. From Fig. \ref{} (a), the circuit represents $H_{cost}$ term, where the first $U$ gate takes the parameters, $\theta=0, \phi=-2J_{ij}\gamma_0$, and $\lambda=0$, and the second $U$ gate takes the parameters, $\theta=0, \phi=-2h_{i}\gamma_0$, and $\lambda=0$. Similarly, in Fig. \ref{qdctc_Fig3b} (b), the U gate after the barrier represents the $H_{\text{mixer}}$ term having the parameters $\theta=2\beta_0, \phi=\pi/2$, and $\lambda=-\pi/2$.

\begin{figure}[]
\centering
\includegraphics[scale=0.3]{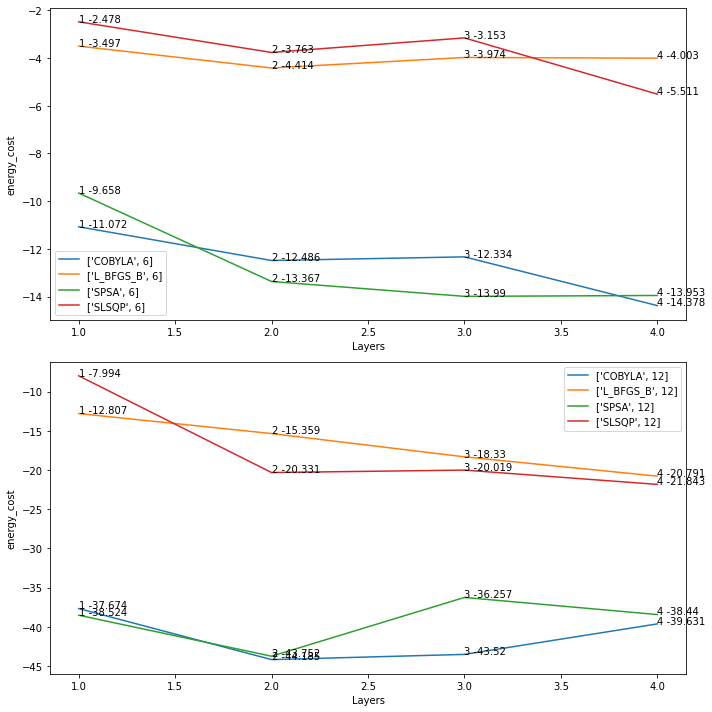}
\caption{Plot illustrating the circuit simulation of VRP with $5$ layers using various optimizers (COBYLA, L\_BFGS\_B, SLSQP, SPSA). The plot consists of two separate graphs depicting the simulation output of $6$ qubit and $12$ qubit circuits, respectively. Each plot, in turn consists of four lines indicating energy values for different optimizers. The average value at each Layer is represented in pairs with (layer, average energy) format.}
\label{VRPCircuitSimulation1}
\end{figure}

\begin{figure}[]
\centering
\includegraphics[scale=0.3]{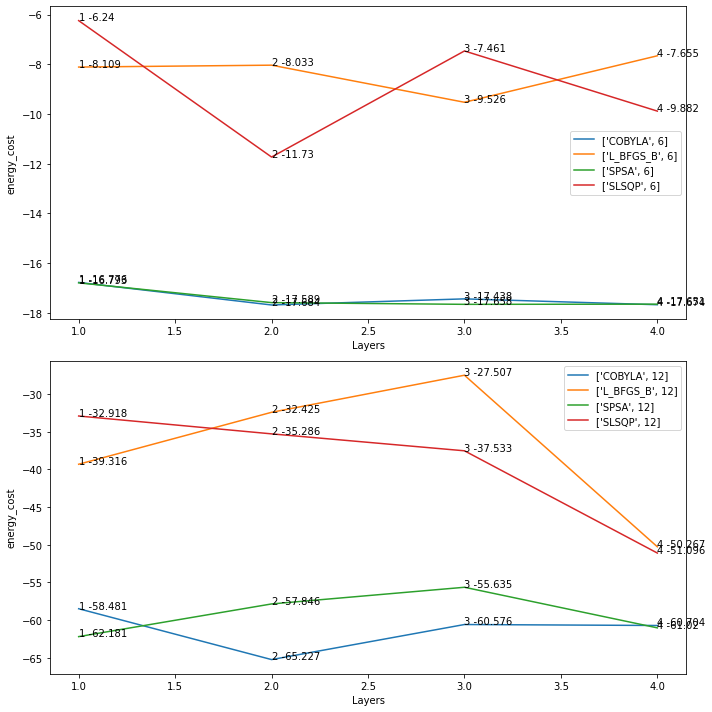}
\caption{Plot illustrating the circuit simulation of VRP with $5$ layers using various optimizers (COBYLA, L\_BFGS\_B, SLSQP, SPSA). The plot consists of two graphs depicting the simulation output of $6$ qubit and $12$ qubit circuits, respectively. Each plot consists of four lines indicating energy values for different optimizers. The minimum value at each Layer is represented in pairs with (layer, minimum energy) format.}
\label{VRPCircuitSimulation2}
\end{figure}

\begin{figure*}[]
\centering
\includegraphics[width = \textwidth]{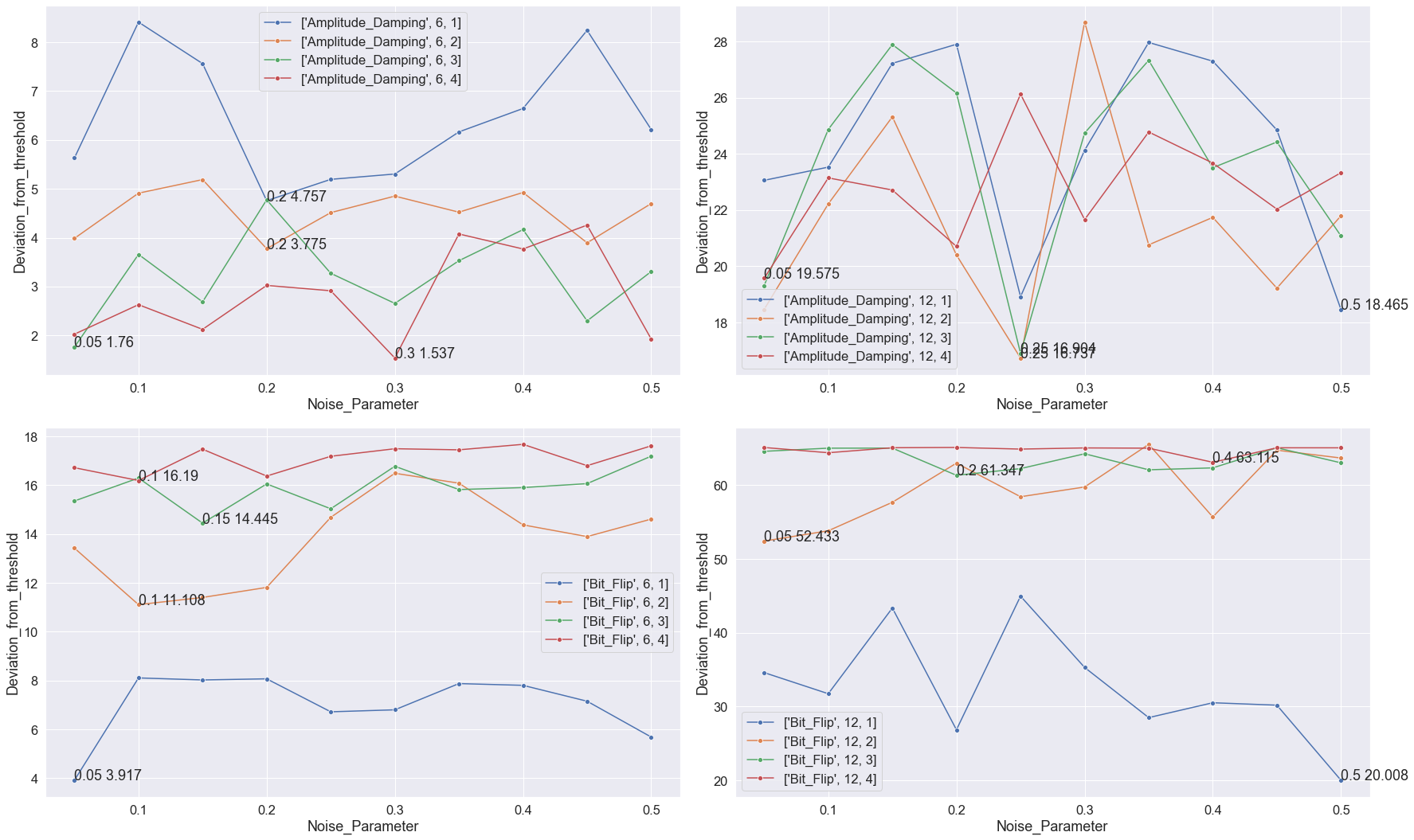}
\caption{Plot illustrating the average Deviation of the energy cost of VRP with $4$ layers using various Amplitude damping and Bit-flip noise models. The plot consists of two charts depicting the simulation output of $6$ qubit and $12$ qubit circuits, respectively. The Average Deviation at each Layer is represented in pairs with (Noise parameter, Average Deviation of Energy cost) format.}
\label{AvarageDeviation AD and BF}
\end{figure*}

\begin{figure*}[]
\centering
\includegraphics[width = \textwidth]{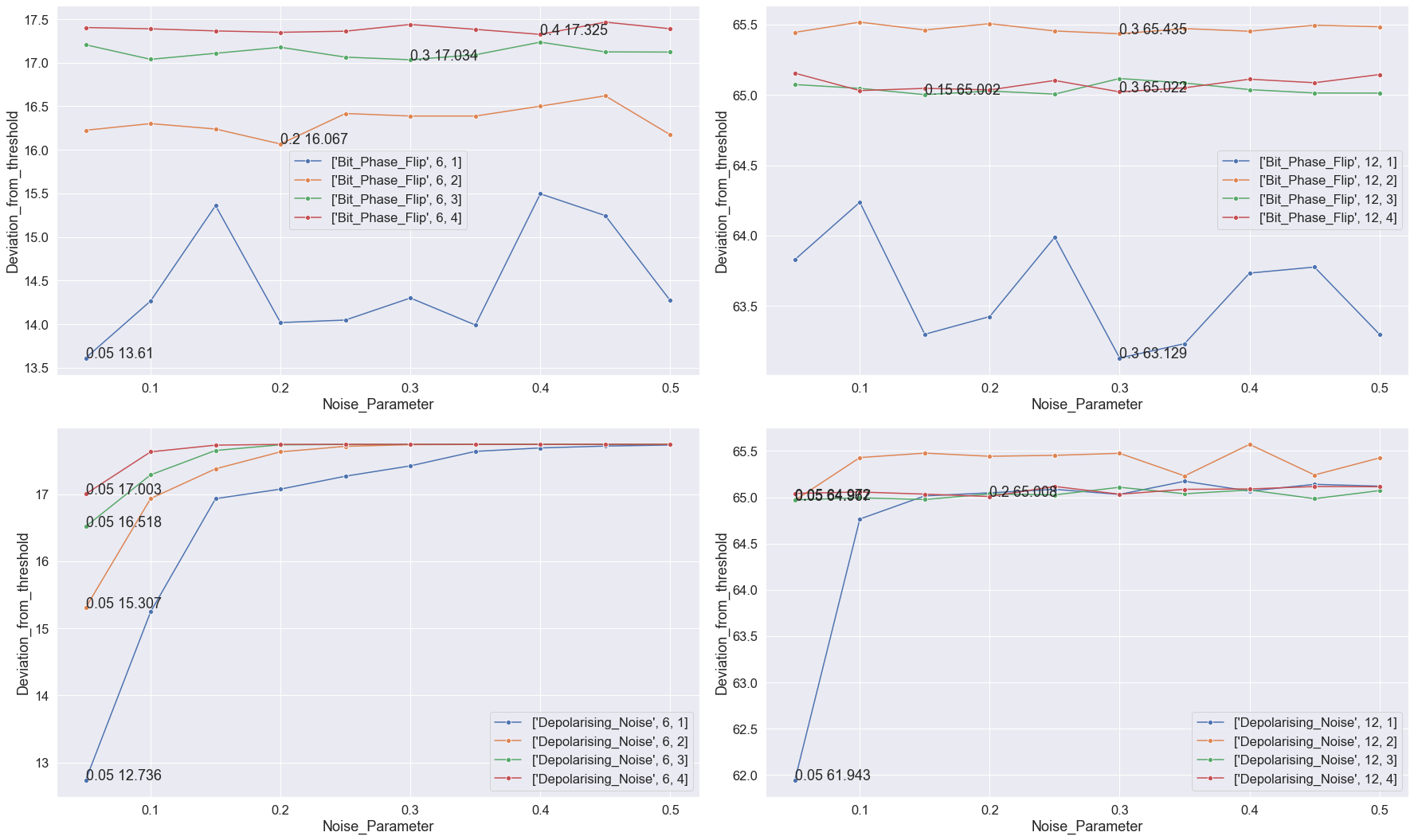}
\caption{Plot illustrating the average Deviation of the energy cost of VRP with $4$ layers using  Bit-phase-flip, and Depolarising noise. The plot consists of two charts depicting the simulation output of $6$ qubit and $12$ qubit circuits, respectively. The Average Deviation at each Layer is represented in pairs with (Noise parameter, Average Deviation of Energy cost) format.}
\label{AvarageDeviation DP and BPF}
\end{figure*}

\begin{figure*}[]
\centering
\includegraphics[width = \textwidth]{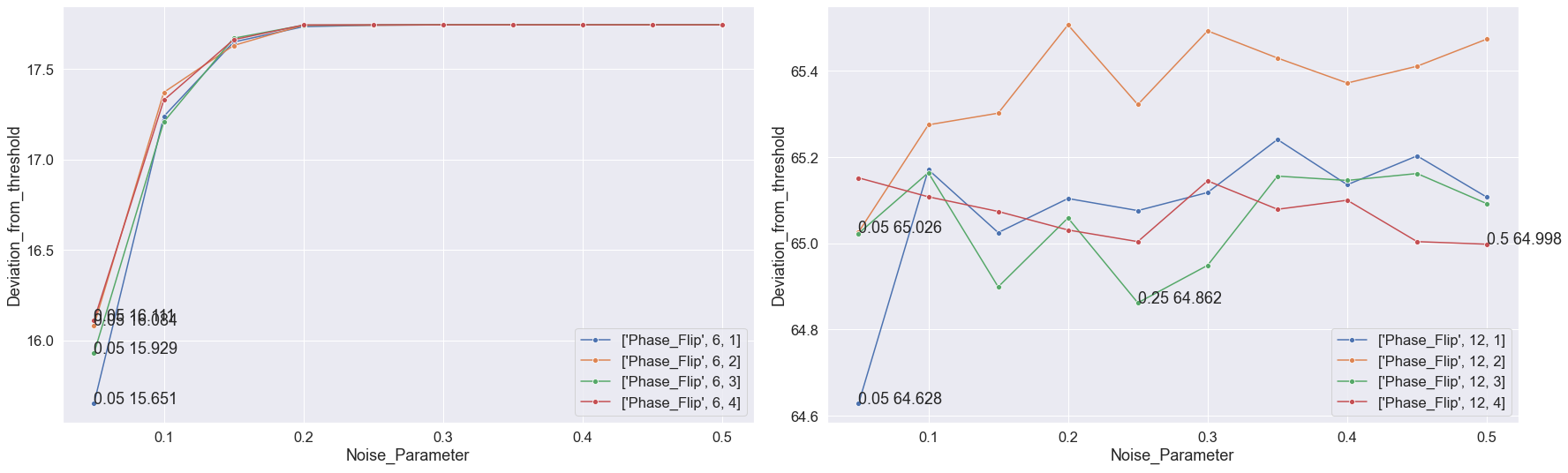}
\caption{Plot illustrating the average energy cost of VRP with $4$ layers using the Phase-flip noise model. The plot consists of two charts depicting the simulation output of $6$ qubit and $12$ qubit circuits, respectively. The average deviation at each Layer is represented in pairs with (Noise parameter, Average Deviation of Energy cost) format.}
\label{AvarageDeviation PF}
\end{figure*}

\begin{table}[!htb]
\centering
\resizebox{\columnwidth}{!}{
\arrayrulecolor{black}
\begin{tabular}{|c|l|l|c|c|c|c|c|} 
\hline
{\cellcolor[rgb]{0.902,0.722,0.718}}                & \multicolumn{1}{c|}{}                & \multicolumn{1}{c|}{}                          & \multicolumn{5}{c|}{\textbf{Noise\_Model}}                                                                                                                                                                                                        \\ 
\hhline{|>{\arrayrulecolor[rgb]{0.902,0.722,0.718}}-~~>{\arrayrulecolor{black}}-----|}
{\cellcolor[rgb]{0.902,0.722,0.718}}\textbf{Qubits} & \multicolumn{1}{c|}{\textbf{Layers}} & \multicolumn{1}{c|}{\textbf{Noise\_Parameter}} & \multicolumn{1}{l|}{\textbf{Amplitude\_Damping}} & \multicolumn{1}{l|}{\textbf{Bit\_Flip}}     & \multicolumn{1}{l|}{\textbf{Bit\_Phase\_Flip}} & \multicolumn{1}{l|}{\textbf{Depolarising\_Noise}} & \multicolumn{1}{l|}{\textbf{Phase\_Flip}}   \\ 
\hline
                                                    & \textbf{1}                           & 0.05                                           & {\cellcolor[rgb]{0.549,0.792,0.49}}-12.053       & {\cellcolor[rgb]{0.482,0.773,0.486}}-13.768 & {\cellcolor[rgb]{0.863,0.882,0.51}}-4.247      & {\cellcolor[rgb]{0.835,0.875,0.506}}-4.949        & {\cellcolor[rgb]{0.953,0.906,0.514}}-2.033  \\ 
\hhline{|~-------|}
                                                    & \textbf{}                            & 0.1                                            & {\cellcolor[rgb]{0.663,0.824,0.498}}-9.273       & {\cellcolor[rgb]{0.651,0.82,0.494}}-9.577   & {\cellcolor[rgb]{0.929,0.902,0.514}}-2.571     & {\cellcolor[rgb]{0.937,0.902,0.514}}-2.437        & {\cellcolor[rgb]{0.988,0.682,0.475}}-0.447  \\
\hline
\end{tabular}}
\caption{VQE simulation of amplitude damping, bit-flip, phase-flip, bit-phase-flip, and depolarizing channel for $6$ and $12$ qubits with $4$ layers involving optimizer COBYLA, where energy costs are averaged over $10$ simulations.}
\label{AvarageEnergyCostTable}
\end{table}

\begin{table}[!htb]
\centering
\resizebox{\columnwidth}{!}{
\arrayrulecolor{black}
\begin{tabular}{|c|c|l|c|c|c|c|c|} 
\hline
{\cellcolor[rgb]{0.902,0.722,0.718}}                &                        & \multicolumn{1}{c|}{}                          & \multicolumn{5}{c|}{\textbf{Noise\_Model}}                                                                                                                                                                                                       \\ 
\hhline{|>{\arrayrulecolor[rgb]{0.902,0.722,0.718}}-~~>{\arrayrulecolor{black}}-----|}
{\cellcolor[rgb]{0.902,0.722,0.718}}\textbf{Qubits} & \textbf{Layers}        & \multicolumn{1}{c|}{\textbf{Noise\_Parameter}} & \multicolumn{1}{l|}{\textbf{Amplitude\_Damping}} & \multicolumn{1}{l|}{\textbf{Bit\_Flip}}    & \multicolumn{1}{l|}{\textbf{Bit\_Phase\_Flip}} & \multicolumn{1}{l|}{\textbf{Depolarising\_Noise}} & \multicolumn{1}{l|}{\textbf{Phase\_Flip}}   \\ 
\hline
                                                    & \textbf{1}             & 0.05                                           & {\cellcolor[rgb]{0.549,0.792,0.49}}5.631         & {\cellcolor[rgb]{0.482,0.773,0.486}}3.917  & {\cellcolor[rgb]{0.863,0.882,0.51}}13.437      & {\cellcolor[rgb]{0.835,0.875,0.506}}12.736        & {\cellcolor[rgb]{0.953,0.906,0.514}}15.651  \\ 
\hhline{|~-------|}
                                                    & \textbf{}              & 0.1                                            & {\cellcolor[rgb]{0.663,0.824,0.498}}8.412        & {\cellcolor[rgb]{0.651,0.82,0.494}}8.107   & {\cellcolor[rgb]{0.929,0.902,0.514}}15.113     & {\cellcolor[rgb]{0.937,0.902,0.514}}15.247        & {\cellcolor[rgb]{0.988,0.682,0.475}}17.238  \\
\hline
\end{tabular}}
\caption{For $6$ and $12$ qubits with $4$ layers and using COBYLA as an optimizer, the table above shows the average deviation from the classical minimum (over $10$ simulations) for VQE simulations utilizing amplitude damping, bit-flip, phase-flip, bit-phase-flip, and depolarising channels.}
\label{DeviationFromThresholdEnergyTable}
\end{table}

\begin{table*}[]
\begin{tabular}{@{}lclllll@{}}
\toprule
\rowcolor[HTML]{2791A6} 
{\color[HTML]{FFFFFF} \textbf{Noise   Channel}} & {\color[HTML]{FFFFFF} \textbf{Qubits}} & {\color[HTML]{FFFFFF} \textbf{Classical   Minimum}} & {\color[HTML]{FFFFFF} \textbf{1st  Layer Dev. \%}} & {\color[HTML]{FFFFFF} \textbf{2nd Layer Dev.\%}} & {\color[HTML]{FFFFFF} \textbf{3rd Layer Dev. \%}} & {\color[HTML]{FFFFFF} \textbf{4th Layer  Dev.\%}} \\ \midrule
\rowcolor[HTML]{CEE7EE} 
\textbf{Amplitude Damping}                      & 6                                      & −17.68                                              & 25\% - 50\%                                        & 25\% - 50\%                                      & 25\% - 50\%                                       & 25\% - 50\%                                       \\
\rowcolor[HTML]{E8F3F7} 
\textbf{Bit Flip}                               & 6                                      & −17.68                                              & 50\%                                               & 75\%                                             & \cellcolor[HTML]{C7A1E3}nearly 100\%              & \cellcolor[HTML]{C7A1E3}nearly 100\%              \\
\rowcolor[HTML]{C7A1E3} 
\cellcolor[HTML]{CEE7EE}\textbf{BitPhase Flip}  & \cellcolor[HTML]{CEE7EE}6              & \cellcolor[HTML]{CEE7EE}−17.68                      & nearly 100\%                                       & nearly 100\%                                     & nearly 100\%                                      & nearly 100\%                                      \\
\rowcolor[HTML]{C7A1E3} 
\cellcolor[HTML]{E8F3F7}\textbf{Phase Flip}     & \cellcolor[HTML]{E8F3F7}6              & \cellcolor[HTML]{E8F3F7}−17.68                      & nearly 100\%                                       & nearly 100\%                                     & nearly 100\%                                      & nearly 100\%                                      \\
\bottomrule
\end{tabular}
\label{Summary of 6qubit Simulation Results Across Noise Models}
\end{table*}

\label{Summary of 6qubit Simulation Results Across Noise Models}

\begin{table*}[]
\begin{tabular}{@{}lclllll@{}}
\toprule
\rowcolor[HTML]{7030A0} 
\multicolumn{1}{c}{\cellcolor[HTML]{7030A0}{\color[HTML]{FFFFFF} \textbf{Noise   Channel}}} & {\color[HTML]{FFFFFF} \textbf{Qubits}} & \multicolumn{1}{c}{\cellcolor[HTML]{7030A0}{\color[HTML]{FFFFFF} \textbf{Classical   Minimum}}} & \multicolumn{1}{c}{\cellcolor[HTML]{7030A0}{\color[HTML]{FFFFFF} \textbf{1st  Layer Dev. \%}}} & \multicolumn{1}{c}{\cellcolor[HTML]{7030A0}{\color[HTML]{FFFFFF} \textbf{2nd Layer   Dev.\%}}} & \multicolumn{1}{c}{\cellcolor[HTML]{7030A0}{\color[HTML]{FFFFFF} \textbf{3rd Layer   Dev. \%}}} & \multicolumn{1}{c}{\cellcolor[HTML]{7030A0}{\color[HTML]{FFFFFF} \textbf{4th   Layer  Dev.\%}}} \\ \midrule
\rowcolor[HTML]{CEE7EE} 
\textbf{Amplitude Damping}                                                                  & 12                                     & −65.684                                                                                        & 25\% - 50\%                                                                                    & 25\% - 50\%                                                                                    & 25\% - 50\%                                                                                     & 25\% - 50\%                                                                                     \\
\rowcolor[HTML]{E8F3F7} 
\textbf{Bit Flip}                                                                           & 12                                     & −65.684                                                                                         & 50\%                                                                                           & 75\%                                                                                           & \cellcolor[HTML]{BF95DF}nearly 100\%                                                            & \cellcolor[HTML]{BF95DF}nearly 100\%                                                            \\
\bottomrule
\end{tabular}
\caption{Table summarizing the inferences on  deviation from the classical minimum for VQE simulation VRP using amplitude damping, bit-flip, phase-flip, bit-phase-flip, and depolarizing channels for $6$ and $12$ qubits.}
\label{Summary of  Simulation Results Across Noise Models}
\end{table*}

\section{VQE Simulation of VRP} \label{VQE Simulation}

We construct the VRP circuit using the above equations and create the Hamiltonians for 3-city and 4-city scenarios. Since we need $n(n-1)$ qubits, we end up with only Hamiltonians and circuits with 6 and 12 qubits. Beyond four cities, it is impossible to simulate in a classical desktop computer due to memory limitations. We create the ansatz using a quantum circuit defined in the previous section and run it across various VQE optimizers available in the IBM Qiskit framework: COBYLA, L\_BFGS\_B, SPSA, and SLSQP. We run the circuit up to $4$ layers across all the optimizers and obtain the results depicted in Fig. \ref{VRPCircuitSimulation1}, and  Fig. \ref{VRPCircuitSimulation2}.

The tables for these figures are presented in Supplementary material section \ref{Supplimentary Figures and Statistics} with tables \ref{VQE Average Energy value of 15 Simulations} and \ref{VQE Minimum Energy value of 15 Simulations}. The figures are derived from 15 consecutive runs of the VRP circuit, each with all the optimizers and four layers. While the average energy values of VRP simulations decrease as optimizations increase across layers for most of the six qubit circuits, the same trend is not observed for 12 qubit circuits. Also, the energy curves vary significantly across all optimizers. Similarly, for minimum energy graphs \ref{VRPCircuitSimulation2}, energy values have no clear decreasing trend as optimization layers increase. Yet from the minimum energy graphs, we can reliably say that the protocol has achieved the minimum energy value, However they do not always follow the downward trend or stay at the same level as optimization layers increase. This, of course, is heavily dependent on the optimizer. Thus when selecting an optimizer for simulation, we chose the optimizer that achieves the lowest minimum, the fewest number of optimization layers. In summary, we have found that COBYLA is the best-performing optimizer, followed by SPSA, L\_BFGS\_B, and SLSQP. However, in the following, when we pass the circuit through various noise models, we will use only the COBYLA optimizer.

\section{Noise Model Simulation of VQE}\label{Noise Model Simulation of VQE}

In a noisy quantum environment, a pure input state will be transformed into a mixed state represented as a density matrix \cite{Marshall_2020,ahadpour_role_2020}. In the case of a 6-qubit state pure state $\ket{\psi}_{q_oq_1q_2q_3q_4q_5}$ the density matrix can be defined as ${\rho}=\ket{\psi}_{q_oq_1q_2q_3q_4q_5}\bra{\psi}_{q_oq_1q_2q_3q_4q_5}$. After the implementation of the noise model, the density matrix takes the following form,
\begin{eqnarray}
{\xi}_{r}(\rho)=\sum_{m}(E_{m}^{rq0})(E_{m}^{rq1})...(E_{m}^{rq5}){\rho}\nonumber\\ 
\times(E_{m}^{xq0})^{\dagger}(E_{m}^{xq1})^{\dagger}...(E_{m}^{xq5})^{\dagger},
\end{eqnarray}
where ${r}\in\{A, B, W, F, D\}$. The elements of the noise channels are described as follows, $A$ is amplitude damping noise, $B$ is bit-flip noise, $W$ is phase-flip noise, $F$ is bit-phase-flip noise, and $D$ is depolarising noise. We apply these noise channels to our VRP circuit and ansatz, which is variable based on the number of qubits ($6$ or $12$) and layers ($1$ to $5$). For simulation purposes, we choose the optimizer COBYLA as it has the best performance characteristics in the simulation of VQE. We restrict the noise probability to $0.5$ as noisy environments beyond this noise level are unlikely and irrelevant in practice. The following subsections discuss the noise channels and operators we experimented on in the VRP circuit.

\subsection{Amplitude Damping}

The energy dissipation is a consequence of the interaction of the quantum system with an amplitude-damping channel. A quantum system gaining or losing energy from or to its environment is described as a change in amplitude rather than phase \cite{ahadpour_role_2020, zhou_asymmetric_2019}. If $\kappa_{A}$ is the probability of gain or loss of amplitude or decoherence rate, the Kraus operators of amplitude damping channel can be described as follows,

\begin{eqnarray}
{E}_0^{A}&=&\left[\begin{matrix}1&0\\0&\sqrt{1-\kappa_A}\\\end{matrix}\right] , \nonumber\\
{E}_1^{A}&=&\sqrt{\kappa}_{A}\left[\begin{matrix}0&1\\0&0\\\end{matrix}\right].
\label{Eq ADNoise}
\end{eqnarray}

\subsection{Bit-Flip Noise}

Random bit-flip errors characterize bit-flip noise \cite{ zhou_asymmetric_2019} with probability ${\kappa}_{B}$. Thus, the Kraus operators of the bit-flip noise channel can be described as,

\begin{eqnarray}
{E}_{0}^{B}=\sqrt{1- \kappa_{B}} I , \nonumber\\
{E}_{1}^{B}= \sqrt\kappa_BX=\sqrt\kappa_B\left[\begin{matrix}0&1\\1&0\end{matrix}\right] . 
\label{EqBitFlipNoise}
\end{eqnarray}

\subsection{Phase Flip Noise}

Phase-flip noise alters the phase parameter of the quantum system without exchange of energy \cite{ahadpour_role_2020,zhou_asymmetric_2019}. The decoherence rate or the phase-flip noise parameter also follows the simple Bernoulli distribution with probability parameter $\kappa_{W}$. thus the Kraus operators of phase-flip noise channel can be defined as Eq. \eqref{Eq PhaseFlipNoise},

\begin{eqnarray}
E_{0}^{W}&=&\sqrt{1-\kappa_{W}} I , \nonumber\\
E_{1}^{W}&=&\sqrt{\kappa_{W}} Z=\sqrt{\kappa_{W}}\left[\begin{array}{cc}
1 & 0 \\
0 & -1
\end{array}\right] .
\label{Eq PhaseFlipNoise}
\end{eqnarray}

\subsection{Bit-Phase Flip Noise}

Bit-phase flip noise channel is characterized by a combination of random bit-flip errors and a change in the quantum system's phase information without energy loss \cite{zhou_asymmetric_2019}. Like other noise channels, the decoherence rate or the combined probability of bit-phase flip error follows the distribution $\kappa_{F}$. The Kraus operator of the bit-phase flip channel could be given by,

\begin{eqnarray}
E_0^F&=&\sqrt{1-\kappa_{F\ }}I ,\nonumber\\
{E}_1^F&=&\sqrt{\kappa_{F}}Y=\sqrt{\kappa_{F}}\left[\begin{matrix}0&-i\\i&0\\\end{matrix}\right] .
\label{Eq Bit-PhaseFlipNoise}
\end{eqnarray}

\subsection{Depolarizing Noise}

A depolarizing noise channel leaves the system untouched or replaces it with a maximally mixed state of $I/d$ for a $d$-level quantum system. The decoherence rate or the depolarization noise probability follows the distribution with parameter $\kappa_{D}$. The Kraus operators are as follows,

\begin{eqnarray}
\begin{aligned}
&E_{0}^{D}=\sqrt{1-\kappa_{D}} I , \\
&E_{1}^{D}=\sqrt{\frac{\kappa_{D}}{3}} X=\sqrt{\frac{\kappa_{D}}{3}}\left[\begin{array}{ll}
0 & 1 \\
1 & 0
\end{array}\right] , \\
&E_{2}^{D}=\sqrt{\frac{\kappa_{D}}{3}} Y=\sqrt{\frac{\kappa_{D}}{3}}\left[\begin{array}{cc}
0 & -i \\
i & 0
\end{array}\right] , \\
&E_{3}^{D}=\sqrt{\frac{\kappa_{D}}{3}} Z=\sqrt{\frac{\kappa_{D}}{3}}\left[\begin{array}{ll}
1 & 0 \\
0 & -1
\end{array}\right]. 
\end{aligned}
\label{Eq DepolarizingNoise}
\end{eqnarray}
It is to be noted that, in all cases, the noise channel is applied locally to each qubit in the circuit.

\section{Inferences from Simulation} \label{Inferences from Simulation}

In the experiment of simulating VRP across various noise channels, we vary the noise probability from $0.05$ to $0.5$ and observe the energy values of VQE. We execute the VRP circuit with $1$ to $4$ layers on our chosen optimizer COBYLA for both $6$ qubit and $12$ qubit configurations. The experiment is repeated ten times for each noise model with different noise realizations. In the same experiment, we calculate the minimum eigenvalue of classical Hamiltonian and record the difference in energy cost after noise induction. 

The state's energy is recorded for each layer from $1$ to $4$ of the QAOA circuit. These values are averaged over the ten iterations to arrive at the average energy cost for each value of the noise parameter and each layer number. The results are summarized in Table \ref{AvarageEnergyCostTable}. The deviation from the optimal value is shown in Table \ref{DeviationFromThresholdEnergyTable}.

We observe that VQE results are impacted due to the induction of noise. In the below subsections, we will describe our observations briefly for each noise model.

\subsection{Amplitude damping Noise}
Amplitude damping noise shows values range between $50\%$ to $75\%$ of classical minimum for both $6$ qubit ($-17.68$) and $12$ qubit ($-65.684$) circuits. There are a few outliers where the algorithm can reach very close to the classical minimum for $6$ qubit circuits. For $12$ qubit circuits, the values are mostly above $50\%$ of classical minimum but never reach classical minimum as close as in $6$ qubit circuits. This trend is seen across multiple layers for amplitude damping channels. It is noticed that the global minimum across layers is observed at the 2nd layer at $-48.947$, but it is very close to the minimum of the 1st layer at $-47.219$. Hence we can infer that increasing layers does not necessarily improve the results for the amplitude-damping channel. Table \ref{VRP Amplitude Damping Average Energy Cost} summarizes the amplitude damping average energy values. Figure \ref{AvarageDeviation AD and BF} refers to the average deviation of energy cost from the classical minimum at each noise parameter across layers. This again confirms that deviation from classical minimum energy cost due to amplitude damping noise remains within $25\%$ to $50\%$, which is recorded in the table \ref{Summary of  Simulation Results Across Noise Models}.  

\subsection{Bit-Flip Noise}
For the bit-flip noise channel, we note that the VQE values are $50\%$ or above the classical minimum for the first layer, but it degrades to around $25\%$ for 2nd layer, falling further on 3rd and finally close to zero on 4th layer. Since this trend is seen in both $6$ qubit and $12$ qubit circuits, we can infer that increasing the number of layers degrades the VQE values for the bit-flip noise channel. We have summarized the bit-flip noise channel average energy values in the table \ref{VRP Bit-Flip Average Energy Cost}. Figure \ref{AvarageDeviation AD and BF} refers to the average deviation of energy cost from the classical minimum at each noise parameter across layers. This again confirms that deviation from classical minimum energy cost due to bit-flip noise remains within the range of  $50\%$ to $75\%$ for the first two Layers before deteriorating further, which is recorded in the table \ref{Summary of  Simulation Results Across Noise Models}. 

\subsection{Bit-Phase-Flip Noise}
There is a similar observation for the bit-phase-flip channel. The VQE values are $25\%$ percent (or above) of the classical minimum for the first layer, but they degrade as the layers increase for $6$ qubit. For $12$ qubit circuits, the VQE values are consistently poor. We have summarized bit-phase-flip noise channel average energy values in the table \ref{VRP Bit-Phase-Flip Average Energy Cost}. Figure \ref{AvarageDeviation DP and BPF} refers to the average deviation of energy cost from the classical minimum at each noise parameter across layers. This again confirms that deviation from classical minimum energy cost due to bit-phase-flip noise remains close to $100\%$; this is recorded in the table \ref{Summary of  Simulation Results Across Noise Models}.  

\subsection{Phase-Flip and Depolarizing Noise Channel}
Finally, for both depolarizing and phase-flip channels, the VQE values remain close to zero for both $6$ qubit and $12$ qubit circuits. It appears that phase-flip and depolarizing noise channels are the most detrimental in VQE circuits (Supplementary Tables \ref{VRP Phase-Flip Average Energy Cost} and \ref{VRP Depolarising Average Energy Cost}). The figures \ref{AvarageDeviation DP and BPF} and \ref{AvarageDeviation PF} refer to the average deviation of energy cost from the classical minimum at each noise parameter across layers. This again confirms that deviation from classical minimum energy cost due to depolarising and phase-flip noise remains close to $100\%$, recorded in the table \ref{Summary of  Simulation Results Across Noise Models}. 

\subsection{Data gathering and Statistics Collection}
In all the simulations, we have used a quantum instance object and a fixed random seed in the Qiskit framework to avoid VQE terminating early and mitigate statistical fluctuations. Hence all the noise models used here are applied to the quantum instance object, which in turn applies noise to the circuit whose parameters are varied by VQE to arrive at a result. We have executed ten iterations of VQE using various noise channels described above. From the results of the ten simulations, we have taken the average energy value of each noise parameter at each layer. Our figure of merit is the difference between the layer's classical minimum and average energy cost. We remind the reader that gate-based simulations are extremely expensive. The results reported here amounted to 219 hours of CPU time on a standard laptop computer using Qiskit's built-in simulators \cite{Qiskit}. While more iterations would improve the variability of the average energy calculations, some clear trends have already been observed.

\section{Discussion and Conclusion }\label{Discussion and Conclusion}

The work we have presented in this paper provides an interesting avenue for evaluating the effect of noise on detailed gate-based simulations of hybrid quantum algorithms for real-world applications. Noise is considered the most problematic aspect of today's intermediate-scale devices, and hence understanding the details of the effects of noise is critical in understanding how to make the most effective use of them. 

In most cases, the effects of noise are minimal at the first layer of the VRP circuits. While additional layers improve upon the results in the noiseless case, the opposite is valid with the induction of noise. Since some noise will always be present in quantum circuits, an empirical finding of our results is that the COBYLA optimizer performs better for VQE circuits compared to the other available optimizers. Yet there is room to study other optimizers, such as SPSA. We also aim to test and compare these results on more significant VRP instances in physical devices, which is beyond the ability to simulate classically. Future work is also needed to analyze more detailed noise models guided by the measured characteristics from the real NISQ devices proposed to solve problems such as VRP.

\section*{Acknowledgment}
The authors are grateful to the IBM Quantum Experience platform and their team for developing the Qiskit platform and providing open access to their simulators for running quantum circuits and performing the experiments reported here \cite{Qiskit}.

\bibliographystyle{IEEEtran}
\bibliography{IEEE}

\begin{thebibliography}{10}
\providecommand{\url}[1]{#1}
\csname url@samestyle\endcsname
\providecommand{\newblock}{\relax}
\providecommand{\bibinfo}[2]{#2}
\providecommand{\BIBentrySTDinterwordspacing}{\spaceskip=0pt\relax}
\providecommand{\BIBentryALTinterwordstretchfactor}{4}
\providecommand{\BIBentryALTinterwordspacing}{\spaceskip=\fontdimen2\font plus
\BIBentryALTinterwordstretchfactor\fontdimen3\font minus
  \fontdimen4\font\relax}
\providecommand{\BIBforeignlanguage}[2]{{%
\expandafter\ifx\csname l@#1\endcsname\relax
\typeout{** WARNING: IEEEtran.bst: No hyphenation pattern has been}%
\typeout{** loaded for the language `#1'. Using the pattern for}%
\typeout{** the default language instead.}%
\else
\language=\csname l@#1\endcsname
\fi
#2}}
\providecommand{\BIBdecl}{\relax}
\BIBdecl

\bibitem{montanaro_quantum_2016}
\BIBentryALTinterwordspacing
A.~Montanaro, ``Quantum algorithms: an overview,'' \emph{npj Quantum
  Information}, vol.~2, no.~1, p. 15023, Jan. 2016. [Online]. Available:
  \url{https://doi.org/10.1038/npjqi.2015.23}
\BIBentrySTDinterwordspacing

\bibitem{jordan_httpsquantumalgorithmzooorg_nodate}
\BIBentryALTinterwordspacing
S.~Jordan, ``https://quantumalgorithmzoo.org/.'' [Online]. Available:
  \url{https://quantumalgorithmzoo.org/\#ONML}
\BIBentrySTDinterwordspacing

\bibitem{farhi_quantum_2014}
\BIBentryALTinterwordspacing
E.~Farhi, J.~Goldstone, and S.~Gutmann, \emph{A Quantum Approximate
  Optimization Algorithm}, Nov. 2014. [Online]. Available:
  \url{https://arxiv.org/abs/1411.4028v1}
\BIBentrySTDinterwordspacing

\bibitem{farhi_quantum_2000}
\BIBentryALTinterwordspacing
E.~Farhi, J.~Goldstone, S.~Gutmann, and M.~Sipser, \emph{Quantum {Computation}
  by {Adiabatic} {Evolution}}, Jan. 2000. [Online]. Available:
  \url{https://arxiv.org/abs/quant-ph/0001106v1}
\BIBentrySTDinterwordspacing

\bibitem{grover_fast_1996}
\BIBentryALTinterwordspacing
L.~K. Grover, \emph{A fast quantum mechanical algorithm for database search},
  May 1996. [Online]. Available: \url{https://arxiv.org/abs/quant-ph/9605043v3}
\BIBentrySTDinterwordspacing

\bibitem{dasari_solving_2020}
\BIBentryALTinterwordspacing
V.~Dasari, M.~S. Im, and L.~Beshaj, \emph{Solving machine learning optimization
  problems using quantum computers}, M.~Blowers, R.~D. Hall, and V.~R. Dasari,
  Eds.\hskip 1em plus 0.5em minus 0.4em\relax Online Only, United States: SPIE,
  apr 2020. [Online]. Available:
  \url{https://www.spiedigitallibrary.org/conference-proceedings-of-spie/11419/2565038/Solving-machine-learning-optimization-problems-using-quantum-computers/10.1117/12.2565038.full}
\BIBentrySTDinterwordspacing

\bibitem{nac_chapter3_quantum_algo}
\BIBentryALTinterwordspacing
E.~National Academies~of Sciences, ``Chapter: 3 {Quantum} {Algorithms} and
  {Applications},'' in \emph{Quantum {Computing}: {Progress} and {Prospects}
  (2019)}. [Online]. Available:
  \url{https://www.nap.edu/catalog/25196/quantum-computing-progress-and-prospects}
\BIBentrySTDinterwordspacing

\bibitem{harwood_formulating_2021}
\BIBentryALTinterwordspacing
S.~Harwood, C.~Gambella, D.~Trenev, A.~Simonetto, D.~Bernal, and D.~Greenberg,
  ``Formulating and {Solving} {Routing} {Problems} on {Quantum} {Computers},''
  \emph{IEEE Transactions on Quantum Engineering}, vol.~2, pp. 1--17, 2021.
  [Online]. Available: \url{https://ieeexplore.ieee.org/document/9314905}
\BIBentrySTDinterwordspacing

\bibitem{srinivasan_efficient_2018}
\BIBentryALTinterwordspacing
K.~Srinivasan, S.~Satyajit, B.~K. Behera, and P.~K. Panigrahi, ``Efficient
  quantum algorithm for solving travelling salesman problem: An ibm quantum
  experience,'' May 2018. [Online]. Available:
  \url{https://arxiv.org/abs/1805.10928v1}
\BIBentrySTDinterwordspacing

\bibitem{feld_hybrid_2019}
\BIBentryALTinterwordspacing
S.~Feld, C.~Roch, T.~Gabor, C.~Seidel, F.~Neukart, I.~Galter, W.~Mauerer, and
  C.~Linnhoff-Popien, \emph{A {Hybrid} {Solution} {Method} for the
  {Capacitated} {Vehicle} {Routing} {Problem} {Using} a {Quantum} {Annealer}},
  2019, vol.~6. [Online]. Available:
  \url{https://www.frontiersin.org/article/10.3389/fict.2019.00013}
\BIBentrySTDinterwordspacing

\bibitem{nazari_reinforcement_2018}
M.~Nazari, A.~Oroojlooy, L.~Snyder, and M.~Takac, \emph{Reinforcement
  {Learning} for {Solving} the {Vehicle} {Routing} {Problem}}.\hskip 1em plus
  0.5em minus 0.4em\relax Curran Associates, Inc., 2018, vol.~31.

\bibitem{utkarsh_solving_2020}
\BIBentryALTinterwordspacing
U.~Azad, B.~K. Behera, E.~A. Ahmed, P.~K. Panigrahi, and A.~Farouk, ``Solving
  {Vehicle} {Routing} {Problem} {Using} {Quantum} {Approximate} {Optimization}
  {Algorithm},'' \emph{IEEE Transactions on Intelligent Transportation
  Systems}, feb 2022. [Online]. Available:
  \url{https://ieeexplore.ieee.org/document/9774961}
\BIBentrySTDinterwordspacing

\bibitem{papalitsas_qubo_2019}
\BIBentryALTinterwordspacing
C.~Papalitsas, T.~Andronikos, K.~Giannakis, G.~Theocharopoulou, and
  S.~Fanarioti, ``A {QUBO} {Model} for the {Traveling} {Salesman} {Problem}
  with {Time} {Windows},'' \emph{Algorithms}, vol.~12, no.~11, 2019. [Online].
  Available: \url{https://www.mdpi.com/1999-4893/12/11/224}
\BIBentrySTDinterwordspacing

\bibitem{glover_quantum_2020}
\BIBentryALTinterwordspacing
F.~Glover, G.~Kochenberger, M.~Ma, and Y.~Du, ``Quantum {Bridge} {Analytics}
  {II}: {QUBO}-{Plus}, network optimization and combinatorial chaining for
  asset exchange,'' \emph{4OR}, vol.~18, no.~4, pp. 387--417, 2020, publisher:
  Springer. [Online]. Available:
  \url{https://ideas.repec.org/a/spr/aqjoor/v18y2020i4d10.1007\_s10288-020-00464-9.html}
\BIBentrySTDinterwordspacing

\bibitem{kochenberger_unconstrained_2014}
\BIBentryALTinterwordspacing
G.~Kochenberger, J.-K. Hao, F.~Glover, M.~Lewis, Z.~Lü, H.~Wang, and Y.~Wang,
  ``The unconstrained binary quadratic programming problem: {A} survey,''
  \emph{Journal of Combinatorial Optimization}, vol.~28, Jul. 2014. [Online].
  Available: \url{https://link.springer.com/article/10.1007/s10878-014-9734-0}
\BIBentrySTDinterwordspacing

\bibitem{irie_quantum_2019}
\BIBentryALTinterwordspacing
H.~Irie, G.~Wongpaisarnsin, M.~Terabe, A.~Miki, and S.~Taguchi, ``Quantum
  {Annealing} of {Vehicle} {Routing} {Problem} with {Time}, {State} and
  {Capacity},'' in \emph{Quantum {Technology} and {Optimization} {Problems}},
  ser. Lecture {Notes} in {Computer} {Science}, S.~Feld and C.~Linnhoff-Popien,
  Eds.\hskip 1em plus 0.5em minus 0.4em\relax Cham: Springer International
  Publishing, 2019, pp. 145--156. [Online]. Available:
  \url{https://link.springer.com/chapter/10.1007/978-3-030-14082-3\_13}
\BIBentrySTDinterwordspacing

\bibitem{crispin_quantum_2013}
\BIBentryALTinterwordspacing
A.~Crispin and A.~Syrichas, ``Quantum {Annealing} {Algorithm} for {Vehicle}
  {Scheduling},'' in \emph{2013 {IEEE} {International} {Conference} on
  {Systems}, {Man}, and {Cybernetics}}, Oct. 2013, pp. 3523--3528, iSSN:
  1062-922X. [Online]. Available:
  \url{https://ieeexplore.ieee.org/document/6722354}
\BIBentrySTDinterwordspacing

\bibitem{fujitsu_annealer_2019}
\BIBentryALTinterwordspacing
F.~E. Office, ``Application of {Digital} {Annealer} for {Faster}
  {Combinatorial} {Optimization},'' \emph{FUJITSU Sci. Tech. J.}, vol.~55,
  no.~2, p.~7, 2019. [Online]. Available:
  \url{https://www.fujitsu.com/global/documents/about/resources/publications/fstj
  \ /archives/vol55-2/paper12.pdf}
\BIBentrySTDinterwordspacing

\bibitem{campos_training_2021}
\BIBentryALTinterwordspacing
E.~Campos, D.~Rabinovich, V.~Akshay, and J.~Biamonte, ``Training saturation in
  layerwise quantum approximate optimization,'' \emph{Phys. Rev. A}, vol. 104,
  no.~3, p. L030401, Sep. 2021, publisher: American Physical Society. [Online].
  Available: \url{https://link.aps.org/doi/10.1103/PhysRevA.104.L030401}
\BIBentrySTDinterwordspacing

\bibitem{wang_noise-induced_2021}
\BIBentryALTinterwordspacing
S.~Wang, E.~Fontana, M.~Cerezo, K.~Sharma, A.~Sone, L.~Cincio, and P.~J. Coles,
  ``Noise-induced barren plateaus in variational quantum algorithms,''
  \emph{Nat Commun}, vol.~12, no.~1, p. 6961, Nov. 2021, number: 1 Publisher:
  Nature Publishing Group. [Online]. Available:
  \url{https://www.nature.com/articles/s41467-021-27045-6}
\BIBentrySTDinterwordspacing

\bibitem{Marshall_2020}
\BIBentryALTinterwordspacing
J.~Marshall, F.~Wudarski, S.~Hadfield, and T.~Hogg, ``Characterizing local
  noise in {QAOA} circuits,'' \emph{{IOP} {SciNotes}}, vol.~1, no.~2, p.
  025208, aug 2020. [Online]. Available:
  \url{https://doi.org/10.1088/2633-1357/abb0d7}
\BIBentrySTDinterwordspacing

\bibitem{lavrijsen_classical_2020}
\BIBentryALTinterwordspacing
W.~Lavrijsen, A.~Tudor, J.~Müller, C.~Iancu, and W.~de~Jong, ``Classical
  {Optimizers} for {Noisy} {Intermediate}-{Scale} {Quantum} {Devices},'' in
  \emph{2020 {IEEE} {International} {Conference} on {Quantum} {Computing} and
  {Engineering} ({QCE})}, Oct. 2020, pp. 267--277. [Online]. Available:
  \url{https://arxiv.org/abs/2004.03004}
\BIBentrySTDinterwordspacing

\bibitem{cerezo_variational_2021}
\BIBentryALTinterwordspacing
M.~Cerezo, A.~Arrasmith, R.~Babbush, S.~C. Benjamin, S.~Endo, K.~Fujii, J.~R.
  McClean, K.~Mitarai, X.~Yuan, L.~Cincio, and P.~J. Coles, ``Variational
  {Quantum} {Algorithms},'' \emph{Nat Rev Phys}, vol.~3, no.~9, pp. 625--644,
  Sep. 2021, arXiv: 2012.09265. [Online]. Available:
  \url{http://arxiv.org/abs/2012.09265}
\BIBentrySTDinterwordspacing

\bibitem{guerrero_solving_2020}
\BIBentryALTinterwordspacing
N.~Guerrero, ``Solving {Combinatorial} {Optimization} {Problems} using the
  {Quantum} {Approximation} {Optimization} {Algorithm},'' \emph{Theses and
  Dissertations}, Mar. 2020. [Online]. Available:
  \url{https://scholar.afit.edu/etd/3263}
\BIBentrySTDinterwordspacing

\bibitem{albash_adiabatic_2018}
\BIBentryALTinterwordspacing
T.~Albash and D.~A. Lidar, ``Adiabatic quantum computation,'' \emph{Rev. Mod.
  Phys.}, vol.~90, no.~1, p. 015002, Jan. 2018, publisher: American Physical
  Society. [Online]. Available:
  \url{https://link.aps.org/doi/10.1103/RevModPhys.90.015002}
\BIBentrySTDinterwordspacing

\bibitem{grant_adiabatic_2020}
\BIBentryALTinterwordspacing
E.~K. Grant and T.~S. Humble, ``Adiabatic {Quantum} {Computing} and {Quantum}
  {Annealing},'' jul 2020, iSBN: 9780190871994. [Online]. Available:
  \url{https://oxfordre.com/physics/view/10.1093/acrefore/9780190871994.001. \
  0001/acrefore-9780190871994-e-32}
\BIBentrySTDinterwordspacing

\bibitem{hoque_quantum_nodate}
\BIBentryALTinterwordspacing
R.~Hoque, ``The quantum approximate optimization algorithm,'' p.~14. [Online].
  Available: \url{https://ryanhoque.github.io/data/QAOA.pdf}
\BIBentrySTDinterwordspacing

\bibitem{sun_adiabatic_2018}
\BIBentryALTinterwordspacing
Y.~Sun, J.-Y. Zhang, M.~S. Byrd, and L.-A. Wu, ``Adiabatic {Quantum}
  {Simulation} {Using} {Trotterization},'' \emph{arXiv:1805.11568 [quant-ph]},
  Jun. 2018, arXiv: 1805.11568. [Online]. Available:
  \url{http://arxiv.org/abs/1805.11568}
\BIBentrySTDinterwordspacing

\bibitem{zhou_quantum_2020}
\BIBentryALTinterwordspacing
L.~Zhou, S.-T. Wang, S.~Choi, H.~Pichler, and M.~D. Lukin, ``Quantum
  {Approximate} {Optimization} {Algorithm}: {Performance}, {Mechanism}, and
  {Implementation} on {Near}-{Term} {Devices},'' \emph{Phys. Rev. X}, vol.~10,
  no.~2, p. 021067, Jun. 2020, publisher: American Physical Society. [Online].
  Available: \url{https://link.aps.org/doi/10.1103/PhysRevX.10.021067}
\BIBentrySTDinterwordspacing

\bibitem{singh_Ising_2020}
\BIBentryALTinterwordspacing
S.~P. Singh, \emph{The {Ising} {Model}: {Brief} {Introduction} and {Its}
  {Application}}.\hskip 1em plus 0.5em minus 0.4em\relax IntechOpen, Feb. 2020,
  publication Title: Solid State Physics - Metastable, Spintronics Materials
  and Mechanics of Deformable Bodies - Recent Progress. [Online]. Available:
  \url{https://www.intechopen.com/chapters/71210}
\BIBentrySTDinterwordspacing

\bibitem{RevModPhys.39.883}
\BIBentryALTinterwordspacing
S.~G. BRUSH, ``History of the lenz-ising model,'' \emph{Rev. Mod. Phys.},
  vol.~39, pp. 883--893, Oct 1967. [Online]. Available:
  \url{https://link.aps.org/doi/10.1103/RevModPhys.39.883}
\BIBentrySTDinterwordspacing

\bibitem{lucas_Ising_2014}
\BIBentryALTinterwordspacing
A.~Lucas, ``Ising formulations of many {NP} problems,'' \emph{Frontiers in
  Physics}, vol.~2, 2014. [Online]. Available:
  \url{https://www.frontiersin.org/article/10.3389/fphy.2014.00005}
\BIBentrySTDinterwordspacing

\bibitem{peruzzo_variational_2014}
\BIBentryALTinterwordspacing
A.~Peruzzo, J.~McClean, P.~Shadbolt, M.-H. Yung, X.-Q. Zhou, P.~J. Love,
  A.~Aspuru-Guzik, and J.~L. O’Brien, ``A variational eigenvalue solver on a
  photonic quantum processor,'' \emph{Nature Communications}, vol.~5, no.~1, p.
  4213, Jul. 2014. [Online]. Available:
  \url{https://doi.org/10.1038/ncomms5213}
\BIBentrySTDinterwordspacing

\bibitem{date_efficiently_2019}
\BIBentryALTinterwordspacing
P.~Date, R.~Patton, C.~Schuman, and T.~Potok, ``Efficiently embedding {QUBO}
  problems on adiabatic quantum computers,'' \emph{Quantum Information
  Processing}, vol.~18, no.~4, p. 117, Mar. 2019. [Online]. Available:
  \url{https://doi.org/10.1007/s11128-019-2236-3}
\BIBentrySTDinterwordspacing

\bibitem{guerreschi_solving_2021}
\BIBentryALTinterwordspacing
G.~G. Guerreschi, ``Solving {Quadratic} {Unconstrained} {Binary} {Optimization}
  with divide-and-conquer and quantum algorithms,'' \emph{arXiv:2101.07813
  [quant-ph]}, Jan. 2021, arXiv: 2101.07813. [Online]. Available:
  \url{http://arxiv.org/abs/2101.07813}
\BIBentrySTDinterwordspacing

\bibitem{Qiskit}
\BIBentryALTinterwordspacing
M.~S.~A. \emph{et al.}, ``Qiskit: An open-source framework for quantum
  computing,'' 2021. [Online]. Available:
  \url{https://github.com/Qiskit/qiskit/tree/0.25.0}
\BIBentrySTDinterwordspacing

\bibitem{ahadpour_role_2020}
\BIBentryALTinterwordspacing
S.~Ahadpour, F.~Mirmasoudi, S.~Ahadpour, and F.~Mirmasoudi, ``The role of noisy
  channels in quantum teleportation,'' \emph{Revista mexicana de física},
  vol.~66, no.~3, pp. 378--387, Jun. 2020, publisher: Sociedad Mexicana de
  Física. [Online]. Available:
  \url{http://www.scielo.org.mx/scielo.php?script=sci\_abstract\&pid=S0035-001X2020000300378\&lng=es\&nrm=iso\&tlng=en}
\BIBentrySTDinterwordspacing

\bibitem{zhou_asymmetric_2019}
\BIBentryALTinterwordspacing
R.-G. Zhou, Y.-N. Zhang, R.~Xu, C.~Qian, and I.~Hou, ``Asymmetric
  {Bidirectional} {Controlled} {Teleportation} by {Using} {Nine}-{Qubit}
  {Entangled} {State} in {Noisy} {Environment},'' \emph{IEEE Access}, vol.~7,
  pp. 75\,247--75\,264, 2019. [Online]. Available:
  \url{https://ieeexplore.ieee.org/document/8726306}
\BIBentrySTDinterwordspacing

\end{thebibliography}



\section{Supplementary Tables and Statistics}\label{Supplimentary Figures and Statistics}

\subsection{No Noise VQE Simulation Stats}
Tables for VRP Simulation without noise \ref{VQE Average Energy value of 15 Simulations}, \ref{VQE Minimum Energy value of 15 Simulations}
\begin{table*}
\centering
\begin{tabular}{|c|c|l|l|l|l|} 
\hline
                &                 & \multicolumn{4}{c|}{{\cellcolor[rgb]{0.773,0.851,0.945}}\textbf{Average of energy\_cost}}                                                                                                                               \\ 
\hhline{|~~----|}
\textbf{Qubits} & \textbf{Layers} & {\cellcolor[rgb]{0.863,0.902,0.945}}\textbf{COBYLA} & {\cellcolor[rgb]{0.863,0.902,0.945}}\textbf{L\_BFGS\_B} & {\cellcolor[rgb]{0.863,0.902,0.945}}\textbf{SLSQP} & {\cellcolor[rgb]{0.863,0.902,0.945}}\textbf{SPSA}  \\ 
\hline
                & 1               & {\cellcolor[rgb]{0.988,0.988,1}}-9.919              & {\cellcolor[rgb]{0.988,0.988,1}}-3.367                  & {\cellcolor[rgb]{0.988,0.988,1}}-2.381             & {\cellcolor[rgb]{0.988,0.988,1}}-8.753             \\ 
\hhline{|~-----|}
                & 2               & {\cellcolor[rgb]{0.816,0.918,0.851}}-10.985         & {\cellcolor[rgb]{0.388,0.745,0.482}}-4.257              & {\cellcolor[rgb]{0.729,0.882,0.776}}-3.692         & {\cellcolor[rgb]{0.553,0.812,0.624}}-11.854        \\ 
\hhline{|~-----|}
                & 3               & {\cellcolor[rgb]{0.741,0.886,0.788}}-11.467         & {\cellcolor[rgb]{0.698,0.871,0.749}}-3.792              & {\cellcolor[rgb]{0.855,0.933,0.886}}-3.037         & {\cellcolor[rgb]{0.416,0.753,0.506}}-12.850        \\ 
\hhline{|~-----|}
6               & 4               & {\cellcolor[rgb]{0.388,0.745,0.482}}-13.711         & {\cellcolor[rgb]{0.604,0.831,0.671}}-3.931              & {\cellcolor[rgb]{0.388,0.745,0.482}}-5.420         & {\cellcolor[rgb]{0.388,0.745,0.482}}-13.057        \\ 
\hline
                & 1               & {\cellcolor[rgb]{1,0.937,0.612}}-30.920             & {\cellcolor[rgb]{1,0.937,0.612}}-9.480                  & {\cellcolor[rgb]{1,0.937,0.612}}-5.941             & {\cellcolor[rgb]{1,0.937,0.612}}-32.290            \\
\hline
\end{tabular}
\caption{Table summarizing the Average Energy Cost (out of 15 runs) of VQE simulation using 5 Optimizers.}
\label{VQE Average Energy value of 15 Simulations}
\end{table*}

\begin{table*}
\centering
\begin{tabular}{|c|l|l|l|l|l|} 
\hline
                & \multicolumn{1}{c|}{}                & \multicolumn{4}{c|}{{\cellcolor[rgb]{0.902,0.722,0.718}}\textbf{Minimum of energy\_cost}}                                                                                                                               \\ 
\hhline{|~~----|}
\textbf{Qubits} & \multicolumn{1}{c|}{\textbf{Layers}} & {\cellcolor[rgb]{0.863,0.902,0.945}}\textbf{COBYLA} & {\cellcolor[rgb]{0.863,0.902,0.945}}\textbf{L\_BFGS\_B} & {\cellcolor[rgb]{0.863,0.902,0.945}}\textbf{SLSQP} & {\cellcolor[rgb]{0.863,0.902,0.945}}\textbf{SPSA}  \\ 
\hline
                & 1                                    & {\cellcolor[rgb]{0.988,0.988,1}}-16.776             & {\cellcolor[rgb]{0.839,0.925,0.871}}-8.109              & {\cellcolor[rgb]{0.988,0.988,1}}-6.240             & {\cellcolor[rgb]{0.988,0.988,1}}-16.793            \\ 
\hhline{|~-----|}
                & 2                                    & {\cellcolor[rgb]{0.388,0.745,0.482}}-17.684         & {\cellcolor[rgb]{0.867,0.937,0.894}}-8.033              & {\cellcolor[rgb]{0.388,0.745,0.482}}-11.730        & {\cellcolor[rgb]{0.435,0.761,0.522}}-17.589        \\ 
\hhline{|~-----|}
                & 3                                    & {\cellcolor[rgb]{0.549,0.808,0.62}}-17.438          & {\cellcolor[rgb]{0.388,0.745,0.482}}-9.526              & {\cellcolor[rgb]{0.851,0.933,0.882}}-7.461         & {\cellcolor[rgb]{0.388,0.745,0.482}}-17.658        \\ 
\hhline{|~-----|}
6               & 4                                    & {\cellcolor[rgb]{0.392,0.745,0.486}}-17.674         & {\cellcolor[rgb]{0.988,0.988,1}}-7.655                  & {\cellcolor[rgb]{0.588,0.824,0.655}}-9.882         & {\cellcolor[rgb]{0.392,0.745,0.486}}-17.651        \\
\hline
\end{tabular}
\caption{Table containing the Minimum Energy value (out of 15 runs) across each layer  of VQE simulation using 5 Optimizers.}
\label{VQE Minimum Energy value of 15 Simulations}
\end{table*}

\subsection{Amplitude Damping Tables}
Tables for Amplitude damping Noise channel containing Average and minimum energy values at various noise parameter over 10 iterations (Tables \ref{VRP Amplitude Damping Average Energy Cost} and \ref{VRP Amplitude Damping Minimum Energy Cost}).
\begin{table*}
\centering
\begin{tabular}{|c|r|r|r|r|r|r|r|r|r|r|r|} 
\hline
                & \multicolumn{1}{c|}{}                & \multicolumn{10}{c|}{\textbf{Average Energy Cost (Amplitude Damping)}}                                                                                                                                                                                                                                                                                                                                                                                                     \\ 
\cline{3-12}
                & \multicolumn{1}{c|}{}                & \multicolumn{10}{c|}{\textbf{Noise Parameter}}                                                                                                                                                                                                                                                                                                                                                                                                                             \\ 
\cline{3-12}
\textbf{Qubits} & \multicolumn{1}{c|}{\textbf{Layers}} & \textbf{0.05}                               & \textbf{0.1}                                & \textbf{0.15}                               & \textbf{0.2}                                & \textbf{0.25}                               & \textbf{0.3}                                & \textbf{0.35}                               & \textbf{0.4}                                & \textbf{0.45}                               & \textbf{0.5}                                 \\
\hline
\end{tabular}
\caption{Table containing the Average Energy values (out of 10 runs) for Amplitude Damping Noise Channel across each layer  of VQE simulation Using COBYLA.}
\label{VRP Amplitude Damping Average Energy Cost}
\end{table*}

\begin{table*}
\centering
\begin{tabular}{|c|c|c|c|c|c|c|c|c|c|c|c|} 
\hline
                &                 & \multicolumn{10}{c|}{{\cellcolor[rgb]{0.902,0.722,0.718}}\textbf{Minumum Energy Cost (Amplitude Damping)}}                                                                                                                                                                                                                                                                                                                                                                                                                                                               \\ 
\hhline{|~~----------|}
                &                 & \multicolumn{10}{c|}{{\cellcolor[rgb]{0.863,0.902,0.945}}\textbf{Noise Parameter}}                                                                                                                                                                                                                                                                                                                                                                                                                                                                                       \\ 
\hhline{|~~----------|}
\textbf{Qubits} & \textbf{Layers} & \multicolumn{1}{l|}{{\cellcolor[rgb]{0.988,0.988,1}}\textbf{0.05}} & \multicolumn{1}{l|}{{\cellcolor[rgb]{0.839,0.925,0.871}}\textbf{0.1}} & \multicolumn{1}{l|}{{\cellcolor[rgb]{0.988,0.988,1}}\textbf{0.15}} & \multicolumn{1}{l|}{{\cellcolor[rgb]{0.988,0.988,1}}\textbf{0.2}} & \multicolumn{1}{l|}{\textbf{0.25}}          & \multicolumn{1}{l|}{\textbf{0.3}}           & \multicolumn{1}{l|}{\textbf{0.35}}          & \multicolumn{1}{l|}{\textbf{0.4}}           & \multicolumn{1}{l|}{\textbf{0.45}}          & \multicolumn{1}{l|}{\textbf{0.5}}            \\
\hline
\end{tabular}
\caption{Table containing the Minimum Energy values  (out of 10 runs) for Amplitude Damping Noise Channel across each layer of VQE simulation Using COBYLA.}
\label{VRP Amplitude Damping Minimum Energy Cost}
\end{table*}

\subsection{Bit-Flip Tables}
Tables for BitFlip Noise channel containing Average and minimum energy values at various noise parameters over 10 iterations (Tables \ref{VRP Bit-Flip Average Energy Cost} and \ref{VRP Bit-Flip Minimum Energy Cost}).
\begin{table*}
\centering
\begin{tabular}{|c|r|r|r|r|r|r|r|r|r|r|r|} 
\hline
                & \multicolumn{1}{c|}{}                & \multicolumn{10}{c|}{\textbf{Average Energy Cost (Bit Flip)}}                                                                                                                                                                                                                                                                                                                                                                                                            \\ 
\cline{3-12}
                & \multicolumn{1}{c|}{}                & \multicolumn{10}{c|}{\textbf{Noise Parameter}}                                                                                                                                                                                                                                                                                                                                                                                                                           \\ 
\cline{3-12}
\textbf{Qubits} & \multicolumn{1}{c|}{\textbf{Layers}} & \textbf{0.05}                               & \textbf{0.1}                                & \textbf{0.15}                              & \textbf{0.2}                                & \textbf{0.25}                              & \textbf{0.3}                                & \textbf{0.35}                               & \textbf{0.4}                                & \textbf{0.45}                               & \textbf{0.5}                                 \\
\hline
\end{tabular}
\caption{Table containing the Average Energy values (out of 10 runs) for BitFlip Noise Channel across each layer  of VQE simulation Using COBYLA.}
\label{VRP Bit-Flip Average Energy Cost}
\end{table*}

\begin{table*}
\centering
\begin{tabular}{|c|c|c|c|c|c|c|c|c|c|c|c|} 
\hline
                &                 & \multicolumn{10}{c|}{{\cellcolor[rgb]{0.902,0.722,0.718}}\textbf{Minimum Energy Cost (Bit Flip)}}                                                                                                                                                                                                                                                                                                                                                                                                                                                                        \\ 
\hhline{|~~----------|}
                &                 & \multicolumn{10}{c|}{{\cellcolor[rgb]{0.863,0.902,0.945}}\textbf{Noise Parameter} Energy}                                                                                                                                                                                                                                                                                                                                                                                                                                                                                \\ 
\hhline{|~~----------|}
\textbf{Qubits} & \textbf{Layers} & \multicolumn{1}{l|}{{\cellcolor[rgb]{0.988,0.988,1}}\textbf{0.05}} & \multicolumn{1}{l|}{{\cellcolor[rgb]{0.839,0.925,0.871}}\textbf{0.1}} & \multicolumn{1}{l|}{{\cellcolor[rgb]{0.988,0.988,1}}\textbf{0.15}} & \multicolumn{1}{l|}{{\cellcolor[rgb]{0.988,0.988,1}}\textbf{0.2}} & \multicolumn{1}{l|}{\textbf{0.25}}          & \multicolumn{1}{l|}{\textbf{0.3}}           & \multicolumn{1}{l|}{\textbf{0.35}}          & \multicolumn{1}{l|}{\textbf{0.4}}           & \multicolumn{1}{l|}{\textbf{0.45}}          & \multicolumn{1}{l|}{\textbf{0.5}}            \\
\hline
\end{tabular}
\caption{Table containing the Minimum Energy values (out of 10 runs) for BitFlip Noise Channel across each layer  of VQE simulation Using COBYLA.}
\label{VRP Bit-Flip Minimum Energy Cost}
\end{table*}

\subsection{Phase-Flip Tables}
Tables for PhaseFlip Noise channel containing Average and minimum energy values at various noise parameters over 10 iterations (Tables \ref{VRP Phase-Flip Average Energy Cost} and \ref{VRP Phase-Flip Minimum Energy Cost}).

\begin{table*}
\centering
\begin{tabular}{|c|r|r|r|r|r|r|r|r|r|r|r|} 
\hline
                & \multicolumn{1}{c|}{}                & \multicolumn{10}{c|}{\textbf{Average Energy Cost (Phase Flip)}}                                                                                                                                                                                                                                                                                                                                                                                                 \\ 
\cline{3-12}
                & \multicolumn{1}{c|}{}                & \multicolumn{10}{c|}{\textbf{Noise Parameter}}                                                                                                                                                                                                                                                                                                                                                                                                                  \\ 
\cline{3-12}
\textbf{Qubits} & \multicolumn{1}{c|}{\textbf{Layers}} & \textbf{0.05}                              & \textbf{0.1}                               & \textbf{0.15}                              & \textbf{0.2}                               & \textbf{0.25}                              & \textbf{0.3}                               & \textbf{0.35}                              & \textbf{0.4}                              & \textbf{0.45}                              & \textbf{0.5}                                \\
\hline
\end{tabular}
\caption{Table containing the Average Energy values (out of 10 runs) for PhaseFlip Noise Channel across each layer  of VQE simulation Using COBYLA.}
\label{VRP Phase-Flip Average Energy Cost}
\caption{Table containing the Minimum Energy values (out of 10 runs) for BitFlip Noise Channel across each layer  of VQE simulation Using COBYLA.}
\label{VRP Bit-Flip Minimum Energy Cost}
\end{table*}

\begin{table*}
\centering
\begin{tabular}{|c|l|c|c|c|c|c|c|c|c|c|c|} 
\hline
                & \multicolumn{1}{c|}{}                & \multicolumn{10}{c|}{{\cellcolor[rgb]{0.902,0.722,0.718}}\textbf{Minimum Energy Cost (Phase Flip)}}                                                                                                                                                                                                                                                                                                                                                                                                                                                                \\ 
\hhline{|~~----------|}
                & \multicolumn{1}{c|}{}                & \multicolumn{10}{c|}{{\cellcolor[rgb]{0.863,0.902,0.945}}\textbf{Noise Parameter}}                                                                                                                                                                                                                                                                                                                                                                                                                                                                                 \\ 
\hhline{|~~----------|}
\textbf{Qubits} & \multicolumn{1}{c|}{\textbf{Layers}} & \multicolumn{1}{l|}{{\cellcolor[rgb]{0.988,0.988,1}}\textbf{0.05}} & \multicolumn{1}{l|}{{\cellcolor[rgb]{0.839,0.925,0.871}}\textbf{0.1}} & \multicolumn{1}{l|}{{\cellcolor[rgb]{0.988,0.988,1}}\textbf{0.15}} & \multicolumn{1}{l|}{{\cellcolor[rgb]{0.988,0.988,1}}\textbf{0.2}} & \multicolumn{1}{l|}{\textbf{0.25}}         & \multicolumn{1}{l|}{\textbf{0.3}}          & \multicolumn{1}{l|}{\textbf{0.35}}         & \multicolumn{1}{l|}{\textbf{0.4}}          & \multicolumn{1}{l|}{\textbf{0.45}}         & \multicolumn{1}{l|}{\textbf{0.5}}           \\
\hline
\end{tabular}
\caption{Table containing the Minimum Energy values (out of 10 runs) for PhaseFlip Noise Channel across each layer  of VQE simulation Using COBYLA.}
\label{VRP Phase-Flip Minimum Energy Cost}
\end{table*}

\subsection{Bit-Phase-Flip Table}
Tables for Bit-PhaseFlip Noise channel containing Average and minimum energy values at various noise parameters over 10 iterations (Tables \ref{VRP Bit-Phase-Flip Average Energy Cost} and \ref{VRP Bit-Phase-Flip Minimum Energy Cost}).
\begin{table*}
\centering
\begin{tabular}{|c|c|l|l|l|l|l|l|l|l|l|l|} 
\hline
                &                 & \multicolumn{10}{c|}{\textbf{Average Energy Cost (Bit-PhaseFlip)}}                                                                                                                                                                                                                                                                                                                                                                                               \\ 
\cline{3-12}
                &                 & \multicolumn{10}{c|}{\textbf{Noise Parameter}}                                                                                                                                                                                                                                                                                                                                                                                                                   \\ 
\cline{3-12}
\textbf{Qubits} & \textbf{Layers} & \multicolumn{1}{r|}{\textbf{0.05}}         & \multicolumn{1}{r|}{\textbf{0.1}}          & \multicolumn{1}{r|}{\textbf{0.15}}         & \multicolumn{1}{r|}{\textbf{0.2}}          & \multicolumn{1}{r|}{\textbf{0.25}}         & \multicolumn{1}{r|}{\textbf{0.3}}          & \multicolumn{1}{r|}{\textbf{0.35}}         & \multicolumn{1}{r|}{\textbf{0.4}}          & \multicolumn{1}{r|}{\textbf{0.45}}         & \multicolumn{1}{r|}{\textbf{0.5}}           \\
\hline
\end{tabular}
\caption{Table containing the Average Energy values (out of 10 runs) for Bit-Phase-Flip Noise Channel across each layer of VQE simulation Using COBYLA.}
\label{VRP Bit-Phase-Flip Average Energy Cost}
\end{table*}

\begin{table*}
\centering
\begin{tabular}{|c|c|c|c|c|c|c|c|c|c|c|c|} 
\hline
                &                 & \multicolumn{10}{c|}{{\cellcolor[rgb]{0.902,0.722,0.718}}\textbf{Minimum Energy Cost (Bit-Phase-Flip)}}                                                                                                                                                                                                                                                                                                                                                                                                                                                            \\ 
\hhline{|~~----------|}
                &                 & \multicolumn{10}{c|}{{\cellcolor[rgb]{0.863,0.902,0.945}}\textbf{Noise Parameter}}                                                                                                                                                                                                                                                                                                                                                                                                                                                                                 \\ 
\hhline{|~~----------|}
\textbf{Qubits} & \textbf{Layers} & \multicolumn{1}{l|}{{\cellcolor[rgb]{0.988,0.988,1}}\textbf{0.05}} & \multicolumn{1}{l|}{{\cellcolor[rgb]{0.839,0.925,0.871}}\textbf{0.1}} & \multicolumn{1}{l|}{{\cellcolor[rgb]{0.988,0.988,1}}\textbf{0.15}} & \multicolumn{1}{l|}{{\cellcolor[rgb]{0.988,0.988,1}}\textbf{0.2}} & \multicolumn{1}{l|}{\textbf{0.25}}         & \multicolumn{1}{l|}{\textbf{0.3}}          & \multicolumn{1}{l|}{\textbf{0.35}}         & \multicolumn{1}{l|}{\textbf{0.4}}          & \multicolumn{1}{l|}{\textbf{0.45}}         & \multicolumn{1}{l|}{\textbf{0.5}}           \\
\hline
\end{tabular}
\caption{Table containing the Minimum Energy values (out of 10 runs) for Bit-PhaseFlip Noise Channel across each layer of VQE simulation Using COBYLA.}
\label{VRP Bit-Phase-Flip Minimum Energy Cost}
\end{table*}

\subsection{Depolarising Table}
Tables for Depolarising Noise channel containing Average and minimum energy values at various noise parameters over 10 iterations (Tables \ref{VRP Depolarising Average Energy Cost} and \ref{VRP Depolarising Minimum Energy Cost}).

\begin{table*}
\centering
\begin{tabular}{|c|r|r|r|r|r|r|r|r|r|r|r|} 
\hline
                & \multicolumn{1}{c|}{}                & \multicolumn{10}{c|}{\textbf{Average Energy Cost (Depolarising)}}                                                                                                                                                                                                                                                                                                                                                                                                \\ 
\cline{3-12}
                & \multicolumn{1}{c|}{}                & \multicolumn{10}{c|}{\textbf{Noise Parameter}}                                                                                                                                                                                                                                                                                                                                                                                                                   \\ 
\cline{3-12}
\textbf{Qubits} & \multicolumn{1}{c|}{\textbf{Layers}} & \textbf{0.05}                              & \textbf{0.1}                               & \textbf{0.15}                              & \textbf{0.2}                               & \textbf{0.25}                              & \textbf{0.3}                               & \textbf{0.35}                              & \textbf{0.4}                               & \textbf{0.45}                              & \textbf{0.5}                                \\
\hline
\end{tabular}
\caption{Table containing the Average Energy values (out of 10 runs) for Depolarising Noise Channel across each layer  of VQE simulation Using COBYLA.}
\label{VRP Depolarising Average Energy Cost}
\end{table*}

\begin{table*}
\centering
\begin{tabular}{|c|l|c|c|c|c|c|c|c|c|c|c|} 
\hline
                & \multicolumn{1}{c|}{}                & \multicolumn{10}{c|}{{\cellcolor[rgb]{0.902,0.722,0.718}}\textbf{Minimum Energy Cost (Depolarising)}}                                                                                                                                                                                                                                                                                                                                                                                                                                                              \\ 
\hhline{|~~----------|}
                & \multicolumn{1}{c|}{}                & \multicolumn{10}{c|}{{\cellcolor[rgb]{0.863,0.902,0.945}}\textbf{Noise Parameter}}                                                                                                                                                                                                                                                                                                                                                                                                                                                                                 \\ 
\hhline{|~~----------|}
\textbf{Qubits} & \multicolumn{1}{c|}{\textbf{Layers}} & \multicolumn{1}{l|}{{\cellcolor[rgb]{0.988,0.988,1}}\textbf{0.05}} & \multicolumn{1}{l|}{{\cellcolor[rgb]{0.839,0.925,0.871}}\textbf{0.1}} & \multicolumn{1}{l|}{{\cellcolor[rgb]{0.988,0.988,1}}\textbf{0.15}} & \multicolumn{1}{l|}{{\cellcolor[rgb]{0.988,0.988,1}}\textbf{0.2}} & \multicolumn{1}{l|}{\textbf{0.25}}         & \multicolumn{1}{l|}{\textbf{0.3}}          & \multicolumn{1}{l|}{\textbf{0.35}}         & \multicolumn{1}{l|}{\textbf{0.4}}          & \multicolumn{1}{l|}{\textbf{0.45}}         & \multicolumn{1}{l|}{\textbf{0.5}}           \\
\hline
\end{tabular}
\caption{Table containing the Minimum Energy values (out of 10 runs) for Depolarising Noise Channel across each layer of VQE simulation Using COBYLA.}
\label{VRP Depolarising Minimum Energy Cost}
\end{table*}

\end{document}